\documentclass[aip,jcp,reprint,citeautoscript,showpacs,floatfix]{revtex4-1}

\usepackage{amssymb}
\usepackage{amsmath}
\usepackage{amsfonts}

\newcommand{\fancy}{\mathcal}
\newcommand{\FI}{\fancy{I}}
\newcommand{\FY}{\fancy{Y}}
\newcommand{\FZ}{\fancy{Z}}
\newcommand{\FK}{\fancy{K}}
\newcommand{\FT}{\fancy{T}}
\newcommand{\FW}{\fancy{W}}

\newcommand{\lbr}{\left(}
\newcommand{\rbr}{\right)}
\newcommand{\lbrs}{\left[}
\newcommand{\rbrs}{\right]}
\newcommand{\lbrc}{\left\{}
\newcommand{\rbrc}{\right\}}

\newcommand{\ten}[1]{\mathsf{#1}}
\renewcommand{\vec}[1]{\boldsymbol{#1}}
\newcommand{\vech}[1]{\vec{\hat{#1}}}
\newcommand{\mat}[1]{\mathbb{#1}}

\newcommand{\beq}{\begin{eqnarray}}
\newcommand{\eeq}{\end{eqnarray}}
\renewcommand{\d}{{\rm{d}}}
\newcommand{\tr}{{{\rm{Tr}}}}

\newcommand{\half}{\frac{1}{2}}
\newcommand{\aeq}{\approx}

\newcommand{\vnabla}{\nabla}

\newcommand{\Dp}[1]{\partial_{{#1}}}

\newcommand{\Da}{\Dp{\alpha}}
\newcommand{\Dm}{\Dp{\mu}}
\newcommand{\Dn}{\Dp{\nu}}
\newcommand{\Dx}{\Dp{x}}

\newcommand{\Df}[1]{\hat{D}_{{#1}}}
\newcommand{\Dfa}{\Df{\alpha}}
\newcommand{\Dfm}{\Df{\mu}}
\newcommand{\Dfn}{\Df{\nu}}

\newcommand{\Rs}{\eta}
\newcommand{\Rv}{\vec{\Rs}}
\newcommand{\Ra}{\Rs_{\alpha}}

\newcommand{\Rm}{\Rs_{\mu}}
\newcommand{\Rn}{\Rs_{\nu}}

\newcommand{\DN}{d_{N}}
\newcommand{\DM}{d_{M}}
\newcommand{\DNa}{d_{N_a}}
\newcommand{\DNj}{d_{N_j}}
\newcommand{\DNl}{d_{N\lambda}}
\newcommand{\Dai}{d_{ai}}
\newcommand{\Daz}{d_{a0}}

\newcommand{\rcite}[1]{Ref. \onlinecite{#1}}

\newcommand{\rcites}[1]{Refs \onlinecite{#1}}

\newcommand{\eref}[1]{Equation~\ref{#1}}
\newcommand{\Eref}[1]{Equation~\ref{#1}}

\newcommand{\sref}[1]{Section~\ref{#1}}
\newcommand{\Sref}[1]{Section~\ref{#1}}
\newcommand{\aref}[1]{Appendix~\ref{#1}}
\newcommand{\Aref}[1]{Appendix~\ref{#1}}

\newcommand{\Ec}{E_{\rm{c}}}

\newcommand{\Exc}{E_{\rm{xc}}}

\newcommand{\chih}{{\hat{\chi}}}

\newcommand{\Omegat}{\tilde{\Omega}}
\newcommand{\Omegab}{\bar{\Omega}}
\newcommand{\comment}[1]{}

\newcommand{\pr}{^{\prime}}

\newcommand{\vx}{\vec{x}}

\newcommand{\vv}{\vec{v}}
\newcommand{\vr}{\vec{r}}
\newcommand{\vrp}{\vec{r}\pr}
\newcommand{\vhr}{\vech{r}}

\newcommand{\vp}{{\vec{p}}}
\newcommand{\vu}{{\vec{u}}}
\newcommand{\vF}{{\vec{F}}}

\newcommand{\mV}{\mat{V}}
\newcommand{\mD}{\mat{D}}
\newcommand{\mT}{\mat{T}}

\newcommand{\vj}{\vec{j}}

\newcommand{\bra}[1]{\left<#1\right|}
\newcommand{\ket}[1]{\left|#1\right>}
\newcommand{\braket}[2]{\left<#1|#2\right>}

\newcommand{\fE}{\fancy{E}}

\newcommand{\KS}{{\rm{KS}}}
\newcommand{\Kin}{{\rm{Kin}}}
\newcommand{\Pot}{{\rm{Pot}}}
\newcommand{\Ext}{{\rm{Ext}}}
\newcommand{\Int}{{\rm{Int}}}

\newcommand{\psin}{\psi_{,\nu}}
\newcommand{\psia}{\psi_{,\alpha}}

\newcommand{\psimn}{\psi_{,\mu\nu}}

\newcommand{\hh}{\hat{h}}
\newcommand{\n}{n^0}
\newcommand{\T}{T^0}
\newcommand{\Tb}{\bar{T}^0}
\newcommand{\tT}{\ten{T}^0}
\newcommand{\tTb}{\bar{\ten{T}}^0}

\newcommand{\rni}{\frac{1}{\Psi^0}}

\newcommand{\Kh}{\hat{K}}
\newcommand{\Rh}{\hat{R}}
\newcommand{\Kht}{\hat{\ten{K}}}
\newcommand{\Rht}{\hat{\ten{R}}}

\newcommand{\Qht}{\hat{\ten{Q}}}

\newcommand{\NBas}{N_{\rm{Bas}}}

\newcommand{\da}{\tilde{d}}
\newcommand{\dn}{\xi}

\newcommand{\vd}{{\vec{\dn}}}
\newcommand{\vda}{{\tilde{\vec{\dn}}}}

\newcommand{\oneel}{^{1e^-}}
\newcommand{\RhOE}{\Rh\oneel}
\newcommand{\RhtOE}{\Rht\oneel}
\newcommand{\dnOE}{\xi\oneel}
\newcommand{\vuOE}{\vu\oneel}
\newcommand{\vdOE}{\vd\oneel}
\newcommand{\vjOE}{\vj\oneel}

\newcommand{\Rhd}{\hat{R}^{(\vd)}}
\newcommand{\Rhdt}{\hat{\ten{R}}^{(\vd)}}

\newcommand{\epsilont}{\tilde{\epsilon}}

\begin{document}
\title{Quantum Continuum Mechanics Made Simple}
\author{Tim Gould}\affiliation{Qld Micro- and Nanotechnology Centre, %
Griffith University, Nathan, Qld 4111, Australia}
\author{Georg Jansen}\affiliation{Fakult\"at f\"ur Chemie, %
Universit\"at Duisburg-Essen, 45117 Essen, Germany}
\author{I. V. Tokatly}\affiliation{ETSF Scientific Development Centre, %
Departamento de F\'isica de Materiales, Universidad del Pa\'is %
Vasco UPV/EHU, Av. Tolosa 72, E-20018 San Sebasti\'an, Spain}
\affiliation{IKERBASQUE, Basque Foundation for Science, E-48011, Bilbao, Spain}
\author{John F. Dobson}\affiliation{Qld Micro- and Nanotechnology Centre, %
Griffith University, Nathan, Qld 4111, Australia}
\begin{abstract}
In this paper we further explore and develop the quantum
continuum mechanics (QCM) of [Tao \emph{et al}, PRL{\bf 103},086401]
with the aim of making it simpler to use in practice.
Our simplifications relate to the non-interacting part of the QCM equations, 
and primarily refer to practical implementations in which the groundstate 
stress tensor is approximated by its Kohn-Sham version.  
We use the simplified approach to directly prove the exactness of QCM
for one-electron systems via an orthonormal formulation.
This proof sheds light on certain 
physical considerations contained in the QCM theory
and their implication on QCM-based approximations. The
one-electron proof then motivates an approximation
to the QCM (exact under certain conditions)
expanded on the wavefunctions of the Kohn-Sham (KS) equations.
Particular attention is paid to the relationships between
transitions from occupied to unoccupied KS orbitals
and their approximations under the QCM. We also demonstrate
the simplified QCM semi-analytically on an example system.
\end{abstract}
\pacs{31.15.E-,31.15.ee,31.15.xg,31.15.ap}
\maketitle

\section{Introduction}

The study of the continuum mechanics (fluid dynamics)
of quantum electron fluids
is almost as old as quantum mechnics itself. The Thomas-Fermi
model and Madelung dynamics\cite{Madelung1926,*Madelung1927}
are two very early examples.
Unfortunately standard approaches to Continuum Mechanics
become too inaccurate or too complex when applied to larger,
real systems such as molecules and atoms,
although a recent development\cite{Nguyen2009} makes
atomic systems more tractable.
Recent work\cite{Tokatly2005-1,*Tokatly2005-2,Tokatly2007,%
Tao2009,Gao2010,Gould2011}
on Continuum Mechanics (QCM) in a moving Lagrangian frame
has led to the development of a sophisticated approach
for describing linear perturbations to many-electron systems
from groundstate properties only. This approach
appears able to bridge the gap between speed and accuracy
required by modern \emph{ab initio} calculations.

The QCM provides an efficient\cite{Gould2011} alternative to full
time-dependent density functional theory
(tdDFT)\cite{HohenbergKohn,*KohnSham,RungeGross,%
[{See page 1898 of }]vdWReview2010,%
Harl2008,*Harl2009,Lebegue2010,Eshuis2010,*Eshuis2011,*Eshuis2012}
calculations. In its general form it can, in principle, be used
to evaluate the transition frequencies and currents of a real many-electron
quantum system with input from the interacting groundstate
one-particle density matrix and two-particle density.
In the formalism presented here we restrict to a more limited form
that takes, as input, groundstate properties obtained
from a Kohn-Sham\cite{HohenbergKohn,*KohnSham} calculation,
and approximates small changes to the
groundstate via a continuum approach. The ability to work with
only groundstate KS properties as input comes at the expense
of having to deal with higher order mixed derivatives (up to four
derivatives with three indices). Initial indications
suggest, however, that it is both tractable and valid in both
model systems\cite{Tao2009,Gao2010},
as well as the difficult and geometrically very different
case of two interacting two-dimensional electron gas layers
\cite{Gould2011}.

Like Madelung dynamics\cite{Madelung1926,*Madelung1927},
the recent QCM approach describes the
behavior of the fluid displacement vector $\vu$ from which the current
and changes to the density can be described. As the
independent-electron density response $\chih_0$ of a system
can be obtained from $\vu$, a ``QCM-dRPA'' correlation
energy functional\cite{Gould2011} has been developed
that uses the direct Random Phase Approximation (dRPA)
but bypasses the need for unoccupied orbitals
by working in the QCM directly via $\vu$.
The QCM-dRPA can be considered a ``third-rung''
functional according to the ``Jacob's Ladder'' classification scheme of
Perdew \emph{et al}\cite{PerdewRungs,*Perdew2005}. The functional
involves calculating the bare response via the QCM scheme and
solving for the interacting response $\chih_{\lambda}$
under the dRPA where additional
interactions are treated at time-dependent Hartree level via
$\chih_{\lambda}=\chih_0+\lambda\chih_0\hat{v}\chih_{\lambda}$.
The response functions $\chih_0$ and $\chih_{\lambda}$ can
then be used to calculate correlation energies.
Here $\lambda$ is the strength of the interactions
and must be integrated over (analytically in some formulations)
 to obtain the kinetic contribution to the correlation energy.

One particular area where it is hoped that the QCM will
prove broadly useful is in the evaluation of van der Waals
physics, where long-range correlation
is important. Here local-density based
techniques\cite{Grimme2006,Tkatchenko2009,Vydrov2009}
and even vdW-adapted approximations like the
vdW-DF\cite{Dion2004,Rydberg2000,Rydberg2003,Langreth2005}
run into difficulties (see \rcite{Gould2011} and
\rcite{Dobson2011-JPCM} for discussion)
because of their use of local or pairwise approximations.

The strong theoretical relationships between the QCM
and the KS-like system it approximates, arising from its
derivation from a formal, moving Lagrange frame\cite{Tokatly2007}, is
somewhat hidden in the current prescription, especially
to those used to working with orbitals and the Schr\"odinger
equation. While relationships can be established (such as sum rules)
between $\vu$ and common groundstate properties of interest,
they do not always come naturally from the original formulation.

Via adaptation and exploration of its theoretical form the QCM
offers a wider scope for further investigation. This is undertaken
in the present work, as follows:

Firstly, in their original form\cite{Tao2009} the QCM equations
were complicated. In the first application to
vdW physics\cite{Gould2011}, a more compact, simpler and more symmetric
but equivalent form was used (equations 3-5 of \rcite{Gould2011}).
The derivation of this simplified form was not given in
\rcite{Gould2011}, and is presented here for the first time
in \sref{sec:QCMForm} along with a discussion of the particular
version of the groundstate stress tensor that makes this simplified
version of the QCM theory possible. In \sref{sec:Response}
we discuss response functions and Adiabatic Connection/
Fluctuation Dissipation theory (ACFD) correlation energies,
filling in some details not elaborated on in \rcite{Gould2011}.

Secondly, the standard displacement vector $\vu$ has some undesirable
properties in Coulomb-localised systems such as atoms and molecules,
including divergence in the tail of the density distribution.
To deal with this,
in \sref{sec:Ortho} we reformulate the QCM in terms of
an ``orthonormal displacement vector'' $\vd=\sqrt{\n}\vu$
[touched on in equation (68) of \rcite{Gao2010}]
where $\n$ is the electron density of the groundstate.
This alternate approach opens the method up to expansion in basis sets
with decaying tails, such as the widely used\cite{SCMOM-1,*SCMOM-2}
Gaussian-Type Orbitals (GTOs) and Slater-Type Orbitals (STOs)
which are only valid in the expansion of decaying functions.

Thirdly, this reformulated QCM is used in \sref{sec:OneEl}
to demonstrate, directly
from the Schr\"odinger equation, that the QCM gives the exact bare,
linear response $\chih_0$ for one-electron systems.
While this relationship is established and demonstrated
in earlier works (see equations (41)-(46) in \rcite{Tokatly2007},
equation (98) and Appendix~D in \rcite{Tao2009}) a proof
direct from the Schr\"odinger equation has not so far appeared.
This relationship is very important, having as a consequence
that properties dependent on $\chih_0$ such as the (dRPA)
asymptotic van der Waals interaction between atoms,
are exactly reproduced by QCM-dRPA theory for one-electron systems
(and two-electron systems with equal groundstate densities
of spin up and down electrons).
The direct proof demonstrates how the QCM relates exactly
to the first-order change of the one-electron orbital/wavefunction.

Fourthly, the one-electron case is used in \sref{sec:ScalarQCM}
to motivate a second, approximate reformulation of the
tensor QCM into a scalar system.
Here the displacement $\vu$ is approximated as the gradient
of a scalar function $s$. This approach simplifies the
QCM equations at the expense of accuracy in general systems
but is exact for one-electron and one-dimensional systems.
By then expanding $\sqrt{\n}s$ on the set of KS orbitals we also
uncover some of the physics of the QCM in the asymptotic tail
regions of the density, where the electrons behave
like a one-electron system.

Finally, in \sref{sec:Harmonic} we illustrate the work on an example system:
a one-dimensional, non-interacting Harmonic potential model.
Here the one-electron case can be solved analytically while
many of the terms in the many-electron case can be
solved for analytically or using near exact quadrature,
minimising numerical error. The scalar approximation
is an exact reformulation of the QCM in this example.

\subsection{Notation}
\label{ssec:Notation}

In this paper we work entirely in atomic units where
$m_{e^-}=\hbar=e^2/(4\pi\epsilon_0)=1$ such that energies are in Hartee and
distances in Bohr radii. We treat energies as frequencies
with the division by $\hbar$ implicit.

Greek sub/super-scripts are used to refer to Cartesian ($x$, $y$ and $z$)
coordinates and are summed over if repeated.
We use the derivative operator notation
$[\Dm f(\vr)]\equiv [\frac{\partial{f(\vr)}}{\partial{r_{\mu}}}]$.

Cartesian tensors are written with sans serif letters (eg. $\ten{T}$),
while Cartesian vectors appear in bold (eg. $\vec{v}$). Their elements
are typically given as $T_{\mu\nu}$ and $v_{\alpha}$ respectively.
The tensor $\ten{T}=\vec{u}\otimes\vec{v}$ has elements
$T_{\mu\nu}=u_{\mu}v_{\nu}$.
More general matrices use double-line Roman letters (eg. $\mat{M}$) and should
be considered square unless otherwise noted.

Operators involve a hat eg. $\hat{O}$. If a derivative appears in an
operator it can be considered to act \emph{entirely} to the right
unless surrounded by \emph{square} brackets, but will act through
other brackets. Thus $[\Da v_{\alpha}]\equiv[\vnabla\cdot\vv]$,
$\Dm f \equiv [\Dm f] + f\Dm$
and $(\Dm f)g \equiv [\Dm f]g + f[\Dm g] + fg\Dm$. Comma-led
subscripts will sometimes be used to represent derivatives
$[\Dm f_{\nu}]\equiv f_{\nu,\mu}$.

In the context of KS orbitals $j$ or $k$ can typically be any orbital
but we reserve $i$ for occupied orbitals only and
$a$ for unoccupied orbitals only such that
$\sum_i\equiv\sum_{i\textrm{ occ}}$ and
$\sum_{ai}\equiv\sum_{i\textrm{ occ}}\sum_{a\textrm{ unocc}}$.

\section{Original vs operator forms of QCM}
\label{sec:QCMForm}

The quantum continuum mechanics (QCM) formalism developed
in previous works\cite{Tokatly2005-1,*Tokatly2005-2,Tokatly2007,%
Tao2009,Gao2010,Gould2011} transforms the problem of calculating
many-electron behaviour in a quantum mechanical system from
an orbital approach to one in which the coordinate system
itself is transformed via a displacement field $\vu(\vr,t)$.
$\vu$ was originally a classical concept, but it can nevertheless
be defined rigorously for a time dependent many-body quantum state
$\ket{\Phi(t)}$ as follows\cite{Dobson1994}:
\begin{align}
\vu(\vr,t)=&\int_{t_0}^{t}\vv(\vr,t')\d t',
&
\vv(\vr,t)=&\vj(\vr,t)/n(\vr,t)
\label{eqn:dispvel}
\\
\vj(\vr,t)=&\bra{\Phi}\hat{\vj}(\vr)\ket{\Phi},
&
n(\vr,t)=&\bra{\Phi}\hat{n}(\vr)\ket{\Phi}
\end{align}
where $\hat{\vj}(\vr)$ and $\hat{n}(\vr)$ are the standard
current and density operators.

We can interpret $\vv(\vr,t)$ as the fluid velocity, and $\vu(\vr,t)$
as the displacement (from $\vr$) at time $t$ of the fluid element that
was at position $\vr$ at time $t_0$. The continuity equation\cite{Dobson1994},
along with \eqref{eqn:dispvel}, implies that, in a linear response situation
around a stationary density $\n(\vr)$, the current and density perturbation
can be found from $\vu$ through
\begin{subequations}
\begin{align}
\vj(\vr,t)=&\Dp{t} \n(\vr)\vu(\vr,t),
\\
\Dp{t}n(\vr,t)=&-\vnabla\cdot\vj(\vr,t)
\end{align}
\label{eqn:jrules}
\end{subequations}
and
\begin{align}
n^1(\vr,t)=&-\vnabla\cdot \n(\vr)\vu(\vr,t).
\label{eqn:deln}
\end{align}

The development of QCM started with work by
Tokatly\cite{Tokatly2005-1,*Tokatly2005-2,Tokatly2007}, who transformed the
Schr\"odinger equation into the Lagrangian coordinate system that
moves with a fluid element. The pressure tensor that determines
the motion of $\vu$ can be obtained in terms of derivatives of
the energy in this frame with respect to $\vu$, derivatives whose
evaluation requires an analysis of the metric tensor $\ten{g}[\vu]$
arising from the transformation from the rest frame to the Lagrangian frame.
The QCM approximation\cite{Tao2009,Gao2010,Pittalis2011} takes the
time dependent many-body wavefunction as a constant in the
Lagrangian frame, corresponding to the fact that much of the
motion of the rest-frame wavefunction is already dealt with via
the motion of the fluid element. In the linear regime one can
then explicitly evaluate the linear pressure tensor and force $\vF^1$,
without the metric tensor appearing explicitly. $\vF^1$ involves
as input only groundstate properties : density $\n(\vr)$,
Kohn-Sham potential $V^{\KS}(\vr)$, stress tensor $\tT(\vr)$,
and pair density $n_2^0(\vr, \vrp)$.
Here we are mainly interested in using $\vu$ to calculate the bare
(Kohn Sham) response, and for this purpose the pair distribution
$n_2^0(\vr,\vrp)$ is not needed.

Restricting to time-periodic perturbations of form
$f(\vr,t)\equiv f(\vr;\omega)e^{i\omega t}$,
Gao \emph{et al}\cite{Gao2010} showed that 
$\vu(\vr;\omega)$ is governed by the elastic equation
\begin{align}
\omega^2\n\vu=&\vF^{1\Ext}+\vF^{1\Pot}+\vF^{1\Kin}+\vF^{1\Int}
\label{eqn:QCMu}
\end{align}
which is exact for one-electron systems at all frequencies.
For many-body systems it is exact in the limit of high frequency.
Here the force terms are: the applied external force
density
$\vF^{1\Ext}(\vr;\omega)=\n(\vr)\nabla V^{1\Ext}(\vr;\omega)$,
the force from the distortion of the groundstate potential
$\vF^{1\Pot}(\vr;\omega)=\n[\vnabla\otimes\vnabla V^{\Ext}(\vr)]
\cdot\vu(\vr;\omega)$,
the force arising from changes to the kinetic energy
$\vF^{1\Kin}(\vr;\omega)=\delta{\FT_2[\vu]}/\delta{\vu(\vr;\omega)}$,
and the force arising from changes to the internal Coulomb interactions
$\vF^{1\Int}(\vr;\omega)=\delta{\FW_2[\vu]}/\delta{\vu(\vr;\omega)}$.
$\FT_2$ and $\FW_2$ are both functionals of $\vu(\vr;\omega)$.

The original papers\cite{Tao2009,Gao2010} establishing the
linearised QCM theory give $\vF^{1\Kin}$ in
equation 14 of \rcite{Tao2009} (equation 53 of \rcite{Gao2010}).
Rearranging order a little this is
\begin{align}
-F^{1\Kin}_{\mu}=&\Da (2\Tb_{\mu\nu}U_{\nu\alpha}+\Tb_{\nu\alpha}\Dm u_{\nu})
\nonumber\\&
+ \frac14 \Dn \left\{
[\Dn\n]\Dm + [\Dm\n]\Dn- \Dm\n\Dn
\right\}\vnabla\cdot\vu
\nonumber\\&
+\frac12 \Dn\left\{
[\nabla^2\n]U_{\mu\nu} -\Dm([\Da\n]U_{\nu\alpha})
\right\} .
\label{eqn:F1KinTok}
\end{align}
Here $U_{\mu\nu}=\half[\Dn u_{\mu}+\Dm u_{\nu}]$ and
they define a `kinetic stress tensor' $\bar{T}_{\mu \nu}$
[equation (17) in \rcite{Gao2010}] via
\begin{align}
\bar{T}_{\mu \nu} (\vr)=&\frac{1}{2} [(\Dm \Dn\pr + \Dm\pr\Dn)
\rho(\vr,\vrp)]_{\vr=\vrp}
\nonumber\\&
- \frac{1}{4} \delta_{\mu \nu}[\nabla^2 n(\vr)].
\label{eqn:TTokGen} 
\end{align}
which is discussed in greater detail below.
$\rho(\vr,\vrp)$ is the one-particle density matrix of the system.
$\tTb$ in \eqref{eqn:F1KinTok} is the groundstate value
of $\bar{\ten{T}}$.

Up to this point we have left the density, density matrix and potential
terms undefined. In the formal theory these can be the exact quantities
of the groundstate system, but in general these quantities are unknown.
In a typical calculation it is likely that these would need to
be calculated in a Kohn-Sham (KS) DFT based approximation, where
we replace $V^{\Ext}$ by the KS potential $V^{\KS}$ in the groundstate,
and replace other quantities by their KS equivalents.

If we define a system with a KS potential $V^{\KS}$
(approximate or otherwise) then the one-electron
Hamiltonian is
\begin{align}
\lbrc -\half\nabla^2 + V^{\KS}(\vr) \rbrc\psi_j(\vr)=&\epsilon_j\psi_j(\vr)
\end{align}
where $\psi_j(\vr)=\braket{\vr}{j}$ is a one-electron orbital
which are orthonormal under
$\int\d\vr \psi_j^*(\vr)\psi_k(\vr)\equiv\braket{j}{k}=\delta_{jk}$
and where $\epsilon_j$ is its Kohn-Sham eigenvalue.

The KS groundstate one-body density matric $\rho^0$
and one-body density $\n$ are
\begin{align}
\rho^0(\vr,\vrp)=&\sum_jf_j\psi_j^*(\vr)\psi_j(\vrp),
\\
\n(\vr)=&\sum_jf_j|\psi_j(\vr)|^2=\rho^0(\vr,\vr)
\end{align}
where $f_j$ is the occupation of orbital $\ket{j}$ defined as
1 for orbitals with $\epsilon_j<\epsilon_F$ and 0 otherwise
where $\epsilon_F$ is the Fermi energy. $\epsilon_F$ should be chosen
to ensure that $\int\n(\vr)\d\vr=N_e$ where $N_e$ is the total
number of electrons.

\subsection{Kinetic stress tensor}

In general, the stress tensor is defined such that its
divergence gives the force per unit volume. In a classical picture,
the kinetic part $\ten{T}^{\Kin}$ of the stress tensor arises because
each fluid element contains a spread of particle velocities,
deviating from the mean value $\Dp{t}\vu(\vr,t)$ (the velocity of the
fluid element). Because there are particles moving
faster and slower than the fluid element, there is a leakage
of particles into nearby fluid elements, and they bring their
momentum with them, resulting in a force.
Unsurprisingly, then, one way to obtain the elements of the
classical stress tensor $T^{\Kin}_{\mu\nu}$ is to form a
second momentum moment of the classical distribution function
$f(\vr,\vp,t)$, multiplied
by two factors of the momentum deviation vector,
$(\vp-m\Dp{t}\vu)_{\mu}(\vp-m\Dp{t}\vu)_{\nu}$ - thus
measuring a mean square spread of momenta. However there is still ambiguity
because strictly only the divergence of $\ten{T}^{\Kin}$ is defined.

As a consequence of the above,
any kinetic stress tensor $\ten{T}^{\Kin}$ must be real symmetric
[$T^{\Kin}_{\mu\nu}(\vr)=T^{\Kin}_{\nu\mu}(\vr)$] and
obey the groundstate force balance condition
$[\Da \ten{T}^{\Kin}_{\alpha\mu}]=-\n[\Dm V^{\Ext}]$.

For the kinetic stress tensor $\ten{T}$
corresponding to a one-electron density matrix 
$\rho(\vr,\vrp)$ in the absence of a current
we choose the following definition,
which we motivate and derive in \aref{app:Kin},
consistently with the qualitative discussion above, and with the use
of the Wigner distribution:
\begin{align}
T^{\Kin}_{\mu \nu} (\vr)=&\frac{1}{2} (\Dm \Dn\pr + \Dm\pr\Dn)
\rho(\vr,\vrp)|_{\vr=\vrp} 
%\nonumber\\&
- \frac{1}{4} \Dm \Dn n(\vr).
\label{eqn:TmnGen}
\end{align}
Inserting the density matrix from the independent-electron
Kohn-Sham groundstate we obtain
\begin{align}
\T_{\mu\nu}=&\Re\sum_jf_j[\Dm\psi_j^*][\Dn\psi_j] - \frac14\Dm\Dn\n
\label{eqn:T0mn}
\\
\equiv&\half\Re\sum_jf_j([\Dm\psi_j^*][\Dn\psi_j]-\psi_j^*[\Dm\Dn\psi_j])
\label{eqn:T0mnA}
\end{align}
where terms are functions of $\vr$ only.

This particular definition seems to allow for the most
compact presentation of the QCM governing equations
[e.g. \eqref{eqn:RHat} and \eqref{eqn:QCMEig}, discussed later],
and is favoured in this work for this reason.
Unless otherwise noted we subsequently restrict ourselves to this form.

Earlier work on the QCM defines the kinetic stress
tensor slightly differently [see \eqref{eqn:TTokGen}].
For an independent-electron Kohn-Sham groundstate this
expression takes the form
%%%end add
\begin{align}
\Tb_{\mu\nu}=\Re\sum_jf_j[\Dm\psi_j^*][\Dn\psi_j]
-\frac14 \delta_{\mu\nu}[\nabla^2\n]
\label{eqn:T0Tok}
\end{align}
where terms depend on $\vr$ only.
The second components in \eqref{eqn:TTokGen} and \eqref{eqn:T0Tok}
differ from ours [compare \eqref{eqn:TmnGen} and \eqref{eqn:T0mn}]. 
%%% end add

\subsection{Linear QCM made simple}

Gould and Dobson\cite{Gould2011} noted without proof that
$\vF^{1\Kin}\equiv-\Kht\vu$, where
\begin{align}
\Kh_{\mu\nu}=&
\Da \T_{\mu\nu} \Da + \Da \T_{\alpha\nu}\Dm + \Dn \T_{\alpha\mu} \Da
%\nonumber\\&
-\frac14 \Dn\Da \n \Da\Dm
\label{eqn:KHat}
\end{align}
is an Hermitian operator. A full derivation of this expression
appears in \aref{app:Compact}.
Since $\vF^{1\Pot}=\n [\vnabla\otimes\vnabla V^{\KS}]\cdot\vu$,
we can write $\vF^{1\Pot}+\vF^{1\Kin}=\Rht\vu$ and thus
\begin{align}
(\omega^2\n - \Rht)\vu =& F^{1\Ext} + F^{1\Int},
\label{eqn:QCMF}
\\
\Rh_{\mu\nu}=&\n V^{\KS}_{,\mu\nu} - \Kh_{\mu\nu}
\label{eqn:RHat}
\end{align}
where all terms but $\omega$ vary with $\vr$.
%%%and this form  changed 21.01.2012 by JFD
The operator $\Rht$
is manifestly Hermitian, and can be shown to be positive definite.
This form of the QCM is easier to deal with in numerical
calculations, and lends itself nicely to expansion on an
auxilliary basis set\cite{Gould2011}.

Equation \eqref{eqn:KHat} leads to the same force
as \eqref{eqn:F1KinTok} but in a much simplified manner.
Much of this simplification comes from the different choice of
kinetic stress tensor, given by
\eqref{eqn:TmnGen} and \eqref{eqn:T0mn} in our work
and by \eqref{eqn:TTokGen} and \eqref{eqn:T0Tok} in
\rcites{Tao2009,Gao2010}, as discussed in \aref{app:Compact}.

Equations \ref{eqn:T0mn}, \ref{eqn:KHat}, \ref{eqn:QCMF} and \ref{eqn:RHat},
form the foundation of the remaining work in this manuscript.
They form the first stage of the `simplification' of the QCM.

\section{Response functions}
\label{sec:Response}

As mentioned in the introduction, one potential application
of the QCM is in the evaluation of response functions, and through
them the groundstate energy of many-electron systems.
This approach was previously investigated\cite{Gould2011} by
some of the authors and found to work well in the
two-dimensional jellium systems studied. Similarly certain
exact properties of the response were previously investigated
in \rcites{Tao2009,Gao2010}. Here we spend some time expanding
on this previous work.

Let us first look at response within the QCM, and then we will investigate
how the QCM response relates to the true KS response.
In \eqref{eqn:QCMu} $\vF^{1\Int}$ takes into account the
electron-electron interaction,
while $\Rht\vu$ deals with kinetic and potential physics.
The dRPA is equivalent to setting
\begin{align}
\vF^{1\Int}(\vr)=&[\Qht\vu](\vr)
\\
\equiv&-\n(\vr)\nabla\int\frac{\d\vrp}{|\vr-\vrp|}
[\vnabla'\cdot\n(\vrp)\vu(\vrp)]
\nonumber\\
\equiv&\int\d\vrp\ten{Q}(\vr,\vrp)\cdot\vu(\vrp)
\label{eqn:FInt}
\end{align}
where
\begin{align}
\ten{Q}(\vr,\vrp)=&
\n(\vr)\n(\vrp)\lbrs\vnabla\otimes\vnabla\pr\frac{1}{|\vr-\vrp|}\rbrs.
\label{eqn:QtDef}
\end{align}

In the absence of an external field $\vF^{1\Ext}=\vec{0}$ we can find
eigen-mode pairs $\Omega_N$ and $\vu_N$ 
(or $\Omega_{N\lambda}$ and $\vu_{N\lambda}$) through solutions of
\begin{align}
\Omega_N^2\n\vu_N=&\Rht\vu_N,
\label{eqn:QCMEig}
\\
\Omega_{N\lambda}^2\n\vu_{N\lambda}=&\Rht\vu_{N\lambda} + \lambda\Qht\vu_{N\lambda}
\label{eqn:QCMEigL}
\end{align}
where \eqref{eqn:QCMEigL} includes the internal interactions $\lambda\Qht$
at coupling strength $\lambda$ while we use the short-hand
$\Omega_N=\Omega_{N0}$ and $\vu_N=\vu_{N0}$
for the non-interacting case.
Because $\Rht$ and $\Qht$ are Hermitian (and in fact
can be shown\cite{Gao2010} to be positive definite)
and $\n$ is symmetric and positive definite, eigensolutions can be found
that obey the orthonormality condition
\begin{align}
\int\d\vr \n\vu_{N\lambda}^*\cdot\vu_{M\lambda}=\delta_{NM}.
\label{eqn:uOrtho}
\end{align}
The set $\{\vu_{N\lambda}\}$ is also guaranteed to be complete
over a finite basis when $\Rht+\lambda\Qht$ is represented
in the same finite basis.
Furthermore the eigen-values $\Omega_{N\lambda}^2$ 
must be positive. Typically we also sort the modes such that
$\Omega_{N+1\lambda}\geq\Omega_{N\lambda}$
where $N\geq 1$ labels the QCM mode with the $N$th lowest energy.
The displacement $\vu_{N\lambda}$ corresponds to a transition density mode
(the meaning will become clearer later) defined as
\begin{align}
\DNl(\vr)=&-\vnabla\cdot \n(\vr)\vu_{N\lambda}(\vr).
\label{eqn:dNQCM}
\end{align}

If the external force density $\vF^{1\Ext}$ is reintroduced we
can expand the solution of \eqref{eqn:QCMu}
at interaction strength $\lambda$
on the basis $\{\vu_{N\lambda}\}$ such that
$\vu=\sum_N c_{N\lambda}\vu_{N\lambda}$. Here
\begin{align}
c_{N\lambda}=&
\frac{\int\d\vr \n(\vr)\vu_{N\lambda}^*(\vr)\cdot\vF^{1\Ext}(\vr)}%
{\Omega_{N\lambda}^2-\omega^2}
\end{align}
when $\vF^{1\Ext}$ is time-periodic with frequency $\omega$.
The change in density \eqref{eqn:deln} can thus be expanded
on \eqref{eqn:dNQCM} as
\begin{align}
n^1_{\lambda}(\vr,\omega)=&\sum_N c_{N\lambda}\DNl(\vr),
\end{align}
where the sum is over all eigen-solutions.

The density response $\chi_{\lambda}$ of a system is defined as the
change in density in response to a $\delta(\vr-\vrp)$ potential at a
frequency $\omega$ with internal interactions at strength $\lambda$.
This corresponds to an external force
$\vF^{1\Ext}=\n\nabla\delta(\vr-\vrp)e^{-i\omega t}$ and internal force
$\vF^{1\Int}=\lambda\Qht\vu(\omega)e^{-i\omega t}$.
Thus the response takes the form
$\chih_{\lambda}(t)=\chih_{\lambda}(\omega)e^{-i\omega t}$ where
$\chi_{\lambda}(\vr,\vrp;\omega)=
-\sum_N\frac{\DNl^*(\vr)\DNl(\vrp)}{\Omega_{N\lambda}^2-\omega^2}$.
Typically it is easier to work with responses at imaginary
frequency $\omega=i\sigma$ such that
\begin{align}
\chi_0(\vr,\vrp;i\sigma)=&
-\Re\sum_N\frac{\DN^*(\vr)\DN(\vrp)}{\sigma^2+\Omega_N^2},
\label{eqn:Chi0}
\\
\chi_{\lambda}(\vr,\vrp;i\sigma)=&
-\Re\sum_N\frac{\DNl^*(\vr)\DNl(\vrp)}{\sigma^2+\Omega_{N\lambda}^2},
\label{eqn:ChiL}
\end{align}
where \eqref{eqn:Chi0} uses the solutions of \eqref{eqn:QCMEig}
to calculate the \emph{bare} ($\lambda=0$) response
while \eqref{eqn:ChiL} uses \eqref{eqn:QCMEigL}
to solve directly for the interacting response. The $\Re$ is unnecessary
as the sum itself can be guaranteed real but may prove
useful in some situations.

While $\chih_{\lambda}$ defined by \eqref{eqn:ChiL} has useful formal
properties its direct evaluation may be numerically difficult
and can be avoided.
Unless otherwise noted we henceforth set $\vF^{1\Int}=\vec{0}$ and deal with
internal interactions (when required) in a less direct,
but precisely equivalent and more computationally convenient manner
(as discussed tangentially
in \rcite{Gould2011} and in \Aref{app:Ec} of this manuscript).

\subsection{Relationships to KS response}

In a Kohn-Sham system with orbitals $\psi_j(\vr)\equiv\braket{\vr}{j}$,
the exact bare response takes the form\cite{RungeGross}
\begin{align}
\chi^{\rm{KS}}_0(\vr,\vrp;i\sigma)=&
-\Re\sum_{ai}\frac{\Dai^*(\vr)\Dai(\vrp)}{\sigma^2+\Omega_{ai}^2}
\label{eqn:Chi0KS}
\end{align}
where $\Dai$ is a normalised transition density
between unoccupied orbital $\ket{a}$ and occupied orbital $\ket{i}$
while $\Omega_{ai}$ is the transition frequency defined by
\begin{subequations}
\begin{align}
\Dai(\vr)=&\sqrt{2\Omega_{ai}}\psi_a^*(\vr)\psi_i(\vr),
\\
\Omega_{ai}=&\epsilon_a-\epsilon_i>0.
\end{align}\label{eqn:Dai}
\end{subequations}
As noted in Sec. \ref{ssec:Notation}, $i$ is summed over
occupied orbitals only and $a$ over unoccupied orbitals only.

There is a transition current density associated with
$\ket{a}$ and $\ket{i}$ which takes the form
\begin{align}
\vj_{ai}(\vr)=&
\frac{1}{2i}[\psi_a(\vr)\nabla\psi_i^*(\vr)-\psi_i^*(\vr)\nabla\psi_a(\vr)]
\end{align}
and where $i\vnabla\cdot\vj_{ai}=\Omega_{ai}\psi_i\psi_a^*=
\sqrt{\Omega_{ai}/2}\Dai$.
Since $\{\vu_N\}$ is complete
(at least within a given finite basis)
and orthonormal under
\eqref{eqn:uOrtho} we can expand $\vj_{ai}=\n
\sum_N[\int\d\vr \vu_N^*\cdot \vj_{ai}]\vu_N$.
Taking the gradient of $\vj_{ai}$ thus provides the
following relationship between the
KS transition densities $\Dai$
and the QCM density modes $\DN$
\begin{align}
\Dai=&\sum_N K_{aiN}\DN,
&
K_{aiN}=&\frac{i\int\d\vr \vu_N^*\cdot \vj_{ai}}{\sqrt{\Omega_{ai}/2}}
\label{eqn:DaiSum}
\end{align}
so that any $\Dai$ can be expanded in $\{\DN\}$.
Unfortunately, since $\{\Dai\}$ is not necessarily complete,
the converse cannot be guaranteed except in the trivial one-electron case.

Certain exact sum rules [equations (81)-(83) further discussed in
appendix~E of \rcite{Gao2010}]
provide some further restrictions on the various
coefficients. Since the f- and third-moment sum rules
are satisfied by the QCM it follows\cite{Tao2009} that
\begin{align}
1=&\sum_{ai} |K_{aiN}|^2,
&
\Omega_N^2=&\sum_{ai} |K_{aiN}|^2 \Omega_{ai}^2
\label{eqn:SumRules}
\end{align}
where $\Omega_{ai}$ are the Kohn-Sham transition frequencies of the system.
For $N\neq M$ we find
\begin{align}
0=&\sum_{ai} K_{aiN}^*K_{aiM}=\sum_{ai} K_{aiN}^*K_{aiM}\Omega_{ai}^2
\end{align}
which come from inserting \eqref{eqn:DaiSum} into \eqref{eqn:Chi0KS}
and comparing the leading two powers of $1/\sigma^2$ with
\eqref{eqn:Chi0}.

The second sum rule in \eqref{eqn:SumRules} makes the relationship
between $\Omega_N$ and the KS transition frequencies clear.
We may also consider $\DN$ to be an approximation to
collections of the transition densities, with errors
hopefully minimised by the sum rules and exact properties
even though no direct expansion exists.
As discussed later these approximations become exact for
one-electron (or two-electrons with equal spin densities
$n_{\uparrow}=n_{\downarrow}$) systems.

It is also worth noting that the lowest QCM transition frequency
$\Omega_1$ can never be less than the transition frequency
between the highest occupied- and lowest unuccopied-
molecular orbital $\Omega_{LH}=\epsilon_L-\epsilon_H$.
In the non-degenerate case the equality follows
if and only if $|K_{LH1}|^2=1$ with all other $K_{ai1}$ zero.
To prove the inequality we note that
$\Omega_{ai}\geq \Omega_{LH}$ and thus
\begin{align}
\Omega_1^2\geq \sum_{ai} |K_{ai1}|^2\Omega_{LH}^2
\geq \Omega_{LH}^2.
\end{align}
If $|K_{LH1}|^2<1$ then $|K_{LH1}|^2=1-\sum_{ai\neq LH}|K_{ai1}|^2$ and
\begin{align*}
\Omega_1^2-\Omega_{LH}^2
=&(|K_{LH1}|^2-1)\Omega_{LH}^2+\sum_{ai\neq LH}|K_{ai1}|^2\Omega_{ai}^2
\\
=&\sum_{ai\neq LH}|K_{ai1}|^2(\Omega_{ai}^2-\Omega_{LH}^2)
>0
\end{align*}
since $|K_{ai1}|^2>0$ and $\Omega_{ai}^2-\Omega_{LH}^2>0$. Thus
the equality only holds if $|K_{ai1}|^2=1$. A direct consequence
of this is that a KS insulator will remain an insulator
under the QCM.

\subsection{Correlation energies made simple}
%%%% Merged this into previous section - may need undoing

From the bare and interacting response functions it is relatively
straightforward to obtain exchange and correlation energies.
In a true KS response formalism this can be obtained via
the occupied and unoccupied orbitals. In the QCM these are replaced
by $\DN$ and $\Omega_N$ and once these have been obtained
the QCM approximation to the correlation energy can be
calculated\cite{Gould2011}.

We define the Coulomb projection matrix $\mat{W}$ with elements
\begin{align}
W_{NM}=&\int\d\vr \vu_M^*(\vr)\cdot[\Qht\vu_N](\vr)
\\
=&\int\d\vr\d\vrp \frac{\DN(\vr)\DM^*(\vrp)}{|\vr-\vrp|}
\label{eqn:WNM}
\end{align}
and $\mat{L}$ with elements $L_{MN}=\delta_{MN}\Omega_N^2$.
Through the working in \Aref{app:Ec} we can show that
the correlation energy is
\begin{align}
\Ec=&\int_0^1\d\lambda \int_0^{\infty}\frac{\d\sigma}{\pi}
\nonumber\\&\times
\half \tr\lbrs \frac{\mat{W}}{\sigma^2+\mat{L}+\lambda\mat{W}}
-\frac{\mat{W}}{\sigma^2+\mat{L}}
\rbrs
\label{eqn:Ec}
\end{align}
or we can use the Furche-like\cite{Furche2008}
integrated form
\begin{align}
\Ec=&\half\sum_N \lbrs \Omegab_N
-\Omega_N\lbr 1 + \frac{W_{NN}}{2\Omega_N^2}\rbr \rbrs
\label{eqn:EcSum}
\end{align}
where $\Omegab_N^2$ are the eigenvalues of $\Rht+\Qht$
or $\mat{L}+\mat{W}$.
As discussed in \Aref{app:Ec} the two diagonalisations
are formally equivalent but
experience in similar techniques suggests that working in the
transition densities of the bare response will allow for
better convergence. In practice diagonalising $\mat{L}+\mat{W}$
is expected to be faster and numerically more reliable and robust.

Using the eigenvalues of $\mat{L}+\mat{W}$ has a further advantage:
we can use a perturbative solution to find the
eigenvalues of $\mat{L}+\mat{W}$ if
$\Omega_N^2\gg \bar{W}_N$ where $\bar{W}_N=\sum_M|W_{NM}|$
(see \Aref{app:Ec} for details).
We define an $N^*$ such that $\Omega_{N}^2\geq K\bar{W}_{N}\forall N>N^*$
where $K$ is sufficiently large. We then solve the
reduced $N^*\times N^*$ eigen-equation $\mat{L}^*+\mat{W}^*$
to obtain $\Omegab^*_{N=1\ldots N^*}$ and calculate
\begin{align}
\Ec\aeq&\half\sum_{N=1}^{N^*} \lbrs \Omegab_N^*
-\Omega_N\lbr 1 + \beta_N\rbr \rbrs
%\nonumber\\&
- \sum_{N>N^*} \frac{\Omega_N\beta_N^2}{4}
\end{align}
where $\beta_N=W_{NN}/(2\Omega_N^2)$.

From the perspective of energy calculations, \eqref{eqn:EcSum}
is the second main stage of simplification of the QCM for
practical purposes. Energies can be calculated through solutions
of the QCM eigen-equation \eqref{eqn:QCMEig} using
the simplified operator \eqref{eqn:KHat} in \eqref{eqn:RHat}.

\section{Orthonormal Displacement}
\label{sec:Ortho}

So far we have investigated, and simplified for practical purposes,
the QCM in its original context as a set of governing equations
for the displacement $\vu$ [see equations \ref{eqn:T0mn}, \ref{eqn:KHat},
\ref{eqn:RHat} and \ref{eqn:QCMEig}]. From these we have derived other
quantities of interest such as response functions and correlation
energies.
In this section we provide a reformulation of these equations
designed to make applications to bound systems like
atoms and molecules more tractable in general.

If we consider the orthonormality condition
$\int\d\vr\n \vu_N^*\cdot\vu_M=\delta_{NM}$ [\eref{eqn:uOrtho}]
on the displacement eigen-modes in a bound system,
we can see that $\vu_N$ may be permitted to grow as $|\vr|\to\infty$
provided $\sqrt{\n}\vu_N$ decreases. In atomic and molecular systems
all valid solutions will, in fact, grow exponentially
due to the asymptotic form of the orbitals. While formally this is
not a great concern, in practise it makes accurate calculation
more difficult in finite systems.

The orthogonality condition suggests that
we can define an orthonormal fluid displacement $\vd=\sqrt{\n}\vu$
that will be a more natural
quantity to use in these systems as it is guaranteed to decrease.
Here the orthonormal eigen-modes are
\begin{align}
\vd_N=\sqrt{\n}\vu_N
\label{eqn:dDef}
\end{align}
where $\int\d\vr \vd_N^*\cdot\vd_M=\delta_{NM}$ 

While $\vu=\sum_N c_N\vu_N$ has a well-defined physical meaning
(the fluid displacement),
$\vd=\sum c_N\vd_N$ is somewhat harder to interpret. However
some insight can be gained if we
define the groundstate quasi-orbital $\Psi^0=\sqrt{\n}$
and its first order change $\Psi^1$. Since
$n^1=2\Psi^0\Psi^1=-\vnabla\cdot\Psi^0\vd$ it is clear that
\begin{align}
2\Psi^1=&-\lbr \vnabla+\frac{[\nabla\Psi^0]}{\Psi^0} \rbr\cdot\vd
\end{align}
and thus $\vd$ is related to the perturbation of
the quasi-orbital. It is also related, via \eqref{eqn:jrules},
to the groundstate properties $n^1$ and $\vj$ through
\begin{align}
\vj=&\Dp{t}\Psi^0\vd
&
n^1=&-\nabla\Psi^0\vd.
\label{eqn:delnd}
\end{align}

We must calculate $\vd_N$ through the QCM equations \eqref{eqn:QCMEig}
which can be rewritten as
\begin{align}
\Omega_N^2\vd_N=&\Rhdt\vd_N,
\label{eqn:QCMEigd}
\\
\Rhdt=&\frac{1}{\Psi^0}\Rht \frac{1}{\Psi^0}
\label{eqn:RHatd}
\end{align}
where $\Rht$ is defined in equation \eqref{eqn:RHat}.
The change in density of a given mode becomes
\begin{align}
\DN(\vr)=&-\Psi^0\lbr \vnabla + \Rv \rbr\cdot\vd_N
\label{eqn:dNAlt}
\end{align}
where
\begin{align}
\Rv=&\frac{[\nabla\n]}{2\n}=\frac{[\nabla\Psi^0]}{\Psi^0}
=\nabla\log\Psi^0
=\half\nabla\log\n
\label{eqn:dDerv}
\end{align}
is the logarithmic gradient of $\Psi^0$.
\Eref{eqn:dNAlt} can be used in \eqref{eqn:WNM} to calculate
the matrix elements of $\mat{W}$ for use in correlation energy
calculations 

Inserting \eqref{eqn:RHat} into \eqref{eqn:RHatd} gives
\begin{align}
\Rhd_{\mu\nu}=& - \rni(\Da \T_{\mu\nu} \Da
+ \Da \T_{\alpha\nu}\Dm + \Dn \T_{\alpha\mu} \Da)\rni
\nonumber\\&
+ V^{\KS}_{,\mu\nu}+ \frac14 \rni\Dn\Da n \Da\Dm \rni.
\label{eqn:RdHat}
\end{align}
Using the derivative operator identity
\begin{align}
\rni\Da - \Da\rni=&\frac{\Ra}{\Psi^0}.
\label{eqn:dn}
\end{align}
and defining $t_{\mu\nu}=\T_{\mu\nu}/\n$,
allows us to convert \eqref{eqn:RdHat} into the following succinct
and symmetric reformulation
\begin{align}
\Rhd_{\mu\nu}=&V^{\KS}_{,\mu\nu}
+\frac14(\Dn + \Rn)(\nabla^2 - S)(\Dm - \Rm).
\nonumber\\&
- (\Da + \Ra)t_{\mu\nu}(\Da - \Ra)
\nonumber\\&
- (\Da + \Ra)t_{\alpha\nu}(\Dm - \Rm)
\nonumber\\&
- (\Dn + \Rn)t_{\alpha\mu}(\Da - \Ra)
\label{eqn:Rhd}
\end{align}
which we can use in \eqref{eqn:QCMEigd}.
Here
\begin{align}
S=&\nabla^2-(\vnabla+\Rv)\cdot(\vnabla-\Rv)
\nonumber\\
=&[\Da\Ra]+\Ra\Ra
=([\half\nabla^2 \n]/\n-\Ra\Ra).
\label{eqn:SDef}
\end{align}

All functions appearing in these expressions depend on
groundstate orbital wavefunctions and their derivatives only.
Expanding in terms of occupied orbitals they are
\begin{align}
t_{\mu\nu}=&
\frac{\Re\half\sum_i ([\Dm\psi_i^*][\Dn\psi_i]-\psi_i^*[\Dm\Dn\psi_i])}
{\sum_i \psi_i^*\psi_i},
\label{eqn:tmnOrb}
\\
\Ra=&\frac{\Re\sum_i \psi_i^*[\Da\psi_i]}{\sum_i \psi_i^*\psi_i},
\label{eqn:RaOrb}
\\
S=&\frac{\Re\sum_i ([\Da\psi_i^*][\Da\psi_i]+\psi_i^*[\nabla^2\psi_i])}%
{\sum_i \psi_i^*\psi_i} - \Ra\Ra.
\label{eqn:SOrb}
\end{align}
Here the force balance equation
$V^{\KS}_{,\mu}=-(\Da+2\Ra) t_{\alpha\mu}$ replaces the usual
$\n V^{\KS}_{,\mu}=-\Da \T_{\alpha\mu}$. In a one-electron system
with occupied orbital $\psi$ and energy $\epsilon_0$
these reduce to
\begin{subequations}
\begin{align}
\Ra=&\frac{\psia}{\psi},
&
t_{\mu\nu}=\half(\Rm\Rn-\frac{\psimn}{\psi}),
\\
S=&2(V^{\KS}-\epsilon_0).
\end{align}
\label{eqn:OneElOrb}
\end{subequations}

It is worth noting that, for molecular systems with Coulomb-like
nuclear potentials the outermost tail is dominated by
one-electron-like behaviour and:
\begin{enumerate}
\item The denominators of $t_{\mu\nu}$, $\Ra$ and $S$
are densities and thus everywhere positive,
\item $t_{\mu\nu}(\vr)\underset{r\to\infty}{\sim} 0$,
\item $|\Rv(\vr)|\underset{r\to\infty}{\sim}|\sqrt{-2\epsilon_H}|$
where $\epsilon_H$ is the KS eigenvalue of the highest
occupied orbital,
\item $S(\vr)\underset{r\to\infty}{\sim} 2(V^{\KS}(\vr) - \epsilon_H)$.
\end{enumerate}
where $\vr$ is the displacement from the center of the highest occupied
orbital and we ignore leading terms that decay exponentially.
In large molecules these expressions may also hold true
closer to nucleii $A$ with $\epsilon_H$ replaced by the
local $\epsilon_H^A$.

To reiterate, the `orthonormal' reformulation of the QCM equations
is introduced to better deal with finite systems such as
atoms and molecules using common methods such as expansion
on GTOs and STOs. Use of the quasi-derivatives $\Da\pm\Ra$
makes evaluation of the normalised operator \eqref{eqn:Rhd}
fairly straightforward.
Practical outputs, such as correlation energies and responses,
can be obtained via the set of
QCM eigen-modes $\vd_N$ which asymptotically
decay in finite systems. These eigen-modes are solutions of
the eigen-equations $\Omega_N^2\vd_N=\Rhdt\vd_N$
[ie. \eqref{eqn:QCMEigd}] using the operator defined in
\eqref{eqn:Rhd} and are normalised via $\int\d\vr\vd_N^*\cdot\vd_N=1$.

We will now proceed in \sref{sec:OneEl} to investigate how
this reformulation applies to one-electron systems, and through this
provide proof that the QCM is exact in such systems. Using the results
of \sref{sec:OneEl} we then motivate an \emph{approximation}
to the QCM in \sref{sec:ScalarQCM}, that remains exact in
one-electron systems.

\section{One-electron systems}
\label{sec:OneEl}

Following equations (41) and (45-46) of \rcite{Tokatly2007},
it is possible to show that the non-linearised QCM formalism
is equivalent to Madelung hydrodynamics\cite{Madelung1926,*Madelung1927}
in a Lagrangian frame for
one electron systems. If all external fields involve gradients of
scalar potentials only, this is directly equivalent\cite{Wallstrom1994} to
finding solutions of the one-electron, time-dependent
Schr\"odinger equation (SE). We note that both one-electron systems,
and two-electron systems with equal densities of spin up and down
are covered.

Here we show this equivalence directly from the SE in the
linear response limit required by the density response $\chih_0$.
This direct proof provides motivation for the approximation 
to the QCM given in the following \sref{sec:ScalarQCM},
by ensuring it is exact in a one-electron system.

We proceed with the proof as follows:
i) we derive the relationship between the perturbed one-electron
wavefunction and its ``orthonormal displacement vector'' $\vdOE$
to show that the latter is entirely determined by the former
(and vice versa up to a trivial phase
via the Runge-Gross theorem\cite{RungeGross});
ii) we show that finding a free-standing solution of the
linear-perturbed SE for a one-electron system is equivalent
to solving an equation of form $\omega^2\vdOE=\RhtOE\vdOE$;
and iii) we show that $\RhtOE=\Rht$ as defined in \eqref{eqn:RHat}.

In any one-electron system we can set
$V=V^{\KS}-\epsilon_0$ (noting that $V^{\KS}=V^{\Ext}$)
to eliminate the energy of the single occupied orbital.
Thus the groundstate Hamiltonian takes the form
\begin{align}
\lbr -\half\nabla^2+V \rbr\psi\equiv\hh\psi=&0
\end{align}
where $\psi$ is the only occupied electron wavefunction
which we make real.
If we apply a small, time-dependent external potential $V^1(t)$
then we can find a new solution $\psi'(t)$ through
\begin{align}
[\hh + V^1(t)]\psi'(t)=-i\Dp{t}\psi'(t).
\label{eqn:SchP}
\end{align}
which will be perturbed only slightly from the groundstate solution.
We can write the perturbed wavefunction via a change to its
magnitude and a rotation of its phase such that
\begin{align}
\psi'(t)=&[\psi+\psi^1(t)]e^{i\phi^1(t)}
\label{eqn:psip}
\end{align}
or, truncating to first order,
$\psi'(t)\aeq \psi+\psi^1(t)+i\psi\phi^1(t)$.
Thus it is sufficient to calculate $\phi^1$ and $\psi^1$ to
fully determine the perturbed solution.

Inserting \eqref{eqn:psip} into \eqref{eqn:SchP} and 
matching real and imaginary components (to linear order) gives
\begin{align}
\psi\Dp{t}\phi^1(t)=&\psi V^1(t) + \hh\psi^1(t),
\\
\Dp{t}\psi^1(t)=&(\nabla\psi)\cdot[\nabla\phi^1(t)]+\half\psi\nabla^2\phi^1(t).
\label{eqn:psi1}
\end{align}
where the second expression relates $\Dp{t}\psi^1$ directly to
$\nabla\phi^1$.
If we then assume a time-periodic external potential
$V^1(t)=V^1e^{i\omega t}$ it follows that
$\psi^1(t)=\psi^1e^{i\omega t}$ and $\phi^1(t)=\phi^1e^{i\omega t}$
and thus $\Dp{t}\equiv i\omega$. We
can then use \eqref{eqn:psi1} to eliminate $\psi^1$ and
derive the equations governing
density perturbations in terms of $\phi^1$ only. Here
\begin{align}
-\omega^2\phi^1=&\half\psi^{-1}\hh(\psia+\Da\psi)\Da\phi^1
+ i\omega V^1(t),
\label{eqn:OneElPhi}
\\
n^1=&2\psi\psi^1=-\frac{\psi}{-i\omega}(\psia+\Da\psi)\Da\phi^1
\label{eqn:OneElDen}
\end{align}
where $n^1$ is the linear change in density and
we have used the derivative relationship
$\Da\psi-\psi\Da\equiv\psia$. By the Runge-Gross
theorem\cite{RungeGross} $V'=V+V^1(t)$ is a functional
of the density $n'=\n+n^1(t)$ only in the linear response regime,
and from \eqref{eqn:OneElDen} it is clear 
that $\psi$ and $\nabla\phi^1$ are sufficient to determine
electronic properties.

The Schr\"odinger current density $\vj$ of the perturbed,
one-electron system is calculated through
\begin{align}
\vjOE=&\frac{1}{2i}
[\psi'\nabla\psi'^*-\psi'^*\nabla\psi']
\aeq-\psi^2 \nabla\phi^1
\end{align}
where we use the first-order perturbation expression
$\psi'\aeq\psi+\psi^1+i\psi\phi^1$ to derive
the second identity. In general\cite{Dobson1994}
the displacement $\vu$ is related to the current via \eqref{eqn:jrules}
and thus $i\omega\vuOE=-\nabla\phi^1$ for the one-electron system.

It then follows trivially that the normalised displacement
$\vdOE=\sqrt{\n}\vuOE\equiv\psi\vuOE$ is
related to $\phi^1$ via
\begin{align}
\vdOE\equiv&\frac{\psi}{-i\omega}\nabla\phi^1
\label{eqn:phi1}
\end{align}
and, from the Runge-Gross theorem\cite{RungeGross}, that $\phi^1$
can be obtained from $\vdOE$. Using \eqref{eqn:OneElDen}
the density perturbation takes the expected form \eqref{eqn:delnd}
$n^1=-\vnabla\cdot\psi\vdOE$.
This completes the first stage of the proof.

By taking the gradient of \eqref{eqn:OneElPhi}
and using $\nabla\phi^1=(-i\omega/\psi)\vdOE$ [from \eqref{eqn:phi1}]
we can explicitly solve the free-standing ($V^1=0$) equation
for $\vdOE$ via
\begin{align}
-\omega^2\frac{\dnOE_{\mu}}{\psi}
=&\half\Dm \psi^{-1}\hh(\psin+\Dn\psi)\frac{\dnOE_{\nu}}{\psi}
\\
\omega^2 \dnOE_{\mu}=&\frac14(\Dm-\Rm)(\nabla^2-S)(\Dn+\Rn)\dnOE_{\nu}
\end{align}
where we used the one-electron specific relationships
\eqref{eqn:OneElOrb} $\Ra=\psia/\psi$ and $2\hh=S-\nabla^2$ for the second
expression. Thus we can define a linear operator
\begin{align}
\RhOE_{\mu\nu}=&\frac14(\Dm-\Rm)(\nabla^2-S)(\Dn+\Rn)
\label{eqn:OEQCM1}
\end{align}
such that $\omega^2 \dnOE_{\mu}=\RhOE_{\mu\nu}\dnOE_{\nu}$ which is
of the same form as \eqref{eqn:QCMu}, as desired for the second
stage of the proof.

Finally it remains to be shown that $\Rht=\RhtOE$.
Following the working in \Aref{app:OneEl} we show that
\begin{align}
\RhOE_{\mu\nu}=&
V_{,\mu\nu} + \frac14(\Dn+\Rn)(\nabla^2-S)(\Dm-\Rm)
\nonumber\\&
-(\Da+\Ra)t_{\mu\nu}(\Da-\Ra)
\nonumber\\&
-(\Da+\Ra) t_{\alpha\nu}(\Dm-\Rm)
\nonumber\\&
-(\Dn+\Rn) t_{\alpha\mu}(\Da-\Ra)
\nonumber\\
\equiv&\Rhd_{\mu\nu}
\label{eqn:RhdComp}
\end{align}
and the proof is complete.

Thus we have shown that i) $\vdOE$ is bijectively (up to a phase)
related to the first-order solution of the perturbed SE;
ii) It obeys $\omega^2\vdOE=\RhtOE\vdOE$ for free-standing modes
and iii) $\RhtOE=\Rht$.
Thus the governing equation is identical in both cases
and it follows that $\vdOE\equiv\vd$.
This confirms that the solutions of the QCM equations
are directly equivalent to the solutions of the
perturbed Schr\"odinger equation in one-electron systems,
and that the QCM is thus exact.

\subsection{Transition modes of one-electron systems}

In a one-electron system,
the density response $\chih_0$ calculated via orbital
transitions \eqref{eqn:Chi0KS} must be the same as that
caclulated via the QCM transitions\eqref{eqn:Chi0}.
Thus
\begin{align}
\chi_0=
\sum_a\frac{\Daz(\vr)\Daz(\vrp)}{\Omega_{a0}^2+\sigma^2}
=&
\sum_N \frac{\DN(\vr)\DN(\vrp)}{\Omega_N^2+\sigma^2}.
\label{eqn:OEChi}
\end{align}
where $a$ is summed over the unoccopied orbitals,
while $\Omega_{a0}=\epsilon_a-\epsilon_0$
and $\Daz=\pm\sqrt{2\Omega_{a0}}\psi\psi_a$.
Since the equality must be true for all $\sigma$
it follows that the individual numerators and denominators
of the sums must be paired and we can choose
an $N_a$ such that $\Omega_{N_a}=\Omega_{a0}$
and $\DNa(\vr)=\Daz(\vr)$.

Equating \eqref{eqn:dNAlt} and \eqref{eqn:Dai} gives
\begin{align}
\DNa=&-\psi(\vnabla+\Rv)\cdot\vd_{N_a}=\sqrt{2\Omega_{a0}}\psi\psi_a.
\label{eqn:QCMEquivOE}
\end{align}
This has a (non-unique) solution
\begin{align}
\vd_{N_a}=&\frac{1}{\sqrt{2\Omega_{a0}}}\lbr\vnabla - \Rv\rbr\psi_a
\label{eqn:QCMSolnOE}
\end{align}
which must therefore be a valid solution of the QCM
equations. The constant pre-factor comes from
$-(\vnabla+\Rv)\cdot(\vnabla-\Rv)\equiv 2(\hh-\epsilon_0)$
in a one-electron system.

Tao \emph{et al}\cite{Tao2009} tested the one-electron exactness
on general $s$-transitions in the Hydrogen atom. We illustrate
\eqref{eqn:QCMEquivOE} and \eqref{eqn:QCMSolnOE}
on the $1s$ to $2p$ transition. Here the $1s$ orbital is occupied with
orbital $\psi_{1s}=e^{-r}/\sqrt{\pi}$ and energy
$\epsilon_0=\epsilon_{1s}=-\half$
while the $2p_z$ orbital has $\psi_{2p_z}=z e^{-r/2}/\sqrt{32\pi}$
and $\epsilon_{2p_z}=-\frac18$. The density is thus $\n=e^{-2r}/\pi$
and $\Rv=-\vhr$.

Using $\nabla z e^{-r/2}=(\vech{z}-\half z\vhr)e^{-r/2}$
we can test \eqref{eqn:QCMEquivOE} and \eqref{eqn:QCMSolnOE}.
With $\Omega_{2p_z-1s}=\frac38$ we find
\begin{align}
\vd_{2p_z-1s}=&\frac{1}{\sqrt{96\pi}} e^{-r/2}(z\vech{r}+2\vech{z}),
\\
d_{2p_z-1s}=&-\frac{\psi_{1s}}{\sqrt{96\pi}}
(\vnabla-\vhr)\cdot (z\vech{r}+2\vech{z})e^{-r/2}
\nonumber\\
=&\sqrt{\frac{3}{4}}\psi_{1s}\psi_{2p_z}.
\label{eqn:OEd1s}
\end{align}
in agreement with \eqref{eqn:QCMEquivOE} and
where $3/4=2\Omega_{1s-2p_z}$ as expected. Note that the usual
displacement vector
$\vu_{2p_z-1s}=\frac{e^{r/2}}{\sqrt{96\pi}}(z\vech{r}+2\vech{z})$
grows exponentially with $r$.

We have thus shown that, in one-electron systems, the QCM is exact.
A one-to-one relationship between the QCM density modes and
KS transition densities [\eref{eqn:QCMSolnOE}] is also established
which we will use as motivation for an approximation.

\section{Scalar approximation to QCM}
\label{sec:ScalarQCM}

While the exact properties of one-electron systems will not hold true
in general, they do suggest a way to approximate the QCM in
such a way that it remains exact for one-electron systems. This
is done by using the form of the one-electron QCM density modes
defined in \eqref{eqn:QCMSolnOE}, to approximately expand the
density modes in general, many-electron systems. Such an approach
would not be expected to provide accurate vector/tensor properties
(such as the tensor response) but might prove acceptable for
scalar properties (such as the scalar density response).

To begin with we note that any well-behaved vector function
can be written as a gradient plus a curl. We may thus set
\begin{align}
\vd_N(\vr)=&\Psi^0\{\nabla s_N(\vr) - \vnabla\times\vv_N(\vr)\}
\end{align}
where $s_N(\vr)$ and $\vv_N(\vr)$ are arbitrary scalar/vector
functions with appropriate asymptotes (we choose our gauge
to ensure $\vnabla\cdot\vv_N=0$).
The form of \eqref{eqn:QCMSolnOE} suggests that we might
approximate our eigen-solutions $\vd_N$ by setting
\begin{align}
\vda_N(\vr)\aeq&\{\vnabla-\Rv(\vr)\}\phi_N(\vr)
=\Psi^0(\vr)\nabla\frac{\phi_N(\vr)}{\Psi^0(\vr)}
\label{eqn:dDefApp}
\end{align}
which is equivalent to making the approximation
$s_N\aeq \phi_N/\Psi^0$ and $\vv_N\aeq \vec{0}$.
This approximation will be exact for one-electron systems and
one-dimensional (1D) systems\footnote{For one-electron this is true
by construction. In the 1D cases
it follows from the fact that in 1D a scalar function
can be written as the derivative of another function
without the curl.} but is not true in general.

The regular QCM eigen-equation \eqref{eqn:QCMEigd} is equivalent to finding
stationary solutions $\delta \fE/\delta \vd=\vec{0}$ of
$\fE[\vd]=\half\int\d\vr \vd^*\cdot\Rht\cdot\vd$ for vectors
$\vd$ satisfying $\int\d\vr\vd^*\cdot\vd=1$. We called
these solutions $\vd_N$ and they can be found through
$\Omega_N^2\vd_N=\Rht\vd_N$ where $\Rht$ is defined in \eqref{eqn:RHat}.
Under the scalar approximation we restrict our solutions to vectors
$\vda$ expressible as $(\vnabla-\Rv)\phi$ which form a connected
subspace of all possible $\vd$.
We thus look for solutions $\vda_N(\phi)$
of $\fE[\vd]$ which are stationary under variation
of $\phi$ ie. $\delta \fE[\vd(\phi)]/\delta\phi=0$
subject to $\int\d\vr\vda(\phi)\cdot\vda(\phi)=1$.

The restricted solutions can be found directly by setting
$\tilde{\fE}[\phi]=-\half
\int\d\vr\phi^*(\vnabla+\Rv)\cdot\Rht\cdot(\vnabla-\Rv)\phi$
and setting the constraint to
$-\int\d\vr \phi^* (\vnabla+\Rv)\cdot(\vnabla-\Rv) \phi=1$.
The general Hermitian eigen-equation for $\phi_N$ thus becomes
\begin{align}
\Omegat_N^2\hat{N}^{\phi}\phi_N=&\hat{R}^{\phi}\phi_N
\label{eqn:QCMEigphi}
\end{align}
with orthogonal solutions normalised
under $\int\d\vr\phi_N^*\hat{N}^{\phi}\phi_M=\delta_{NM}$. Here
\begin{align}
\hat{N}^{\phi}=&-(\vnabla+\Rv)\cdot(\vnabla-\Rv)=S-\nabla^2
\\
\hat{R}^{\phi}=&-(\Dm+\Rm)\hat{R}_{\mu\nu}(\Dn-\Rn).
\end{align}
It is obvious that $\Omegat_1^2\geq\Omega_1^2$ since
the subspace minimum of $\fE$ must be equal to or
higher than its true minimum.

Such an approximation loses some accuracy and some nice properties
of the true QCM but reduces the problem from a tensor
to a scalar. Its exactness in a variety of systems including
the one-electron case in any number of dimensions
suggests that it might be appropriate for vdW calculations in
molecular systems. Here the vdW physics are often dominated by the
asymptotic regions which show one-electron-like behaviour
(see \Sref{ssec:OneElAs} for further details).

In the remaining subsections we will investigate some of the
practical results of this approximation. These are not intended
to be a thorough investigation of the method, but to provide
some guidance to the numerical and theoretical analysis thereof.

\subsection{Scalar approximation in KS orbitals}
We can expand $\phi_N=\sum_j p_{Nj}\psi_j$ in the KS orbitals
(or any other complete and orthonormal basis set) so that,
from \eqref{eqn:dDefApp},
\begin{align}
\vda_N(\vr)=&\sum_j p_{Nj}\{\vnabla-\Rv(\vr)\}\psi_j(\vr)
\label{eqn:dDefKS}
\end{align}
where $\psi_j$ are KS orbitals (occupied or otherwise). In reality
we must truncate to the lowest $\NBas$ orbitals.
The transition density modes are thus (remembering that $\sqrt{\n}=\Psi^0$)
\begin{subequations}
\begin{align}
\da_N(\vr)
=&-\Psi^0\sum_jp_{Nj}(\vnabla+\Rv)\cdot(\vnabla-\Rv)\psi_j
\\
=&\Psi^0\sum_jp_{Nj}(S-\nabla^2)\psi_j
\\
=&\Psi^0\sum_jp_{Nj}[2(\epsilon_j-V^{\KS})+S]\psi_j.
\end{align}\label{eqn:dNKS}%
\end{subequations}
where the first two properties are true for any basis set
but the third property is only true for the KS orbitals.

Projection into the KS orbitals allows some insight into
the physical meaning of this approximation to be obtained
by considering the quasi-orbital $\Psi^0$.
Here we can set $\vda_N=(2i/\Psi^0)\sum_j p_{Nj}\vj_j$
where the quasi-transition current
\begin{align}
\vj_j=&\frac{1}{2i}[\Psi^0\lbr\vnabla-\Rv\rbr\psi_j]
=\frac{1}{2i}[\Psi^0\nabla\psi_j-\psi_j\nabla\Psi^0]
\label{eqn:JQuasi}
\end{align}
has a similar form to a transition current
$\vj_{ai}=\frac{1}{2i}(\psi_i\nabla\psi_a-\psi_a\nabla\psi_i)$
with the occupied orbital replaced by the quasi-orbital
of the \emph{total} density.

If we pre-multiply \eqref{eqn:QCMEigd} by $\vda_M$ and integrate
we find $\Omegat_N^2\int\d\vr \vda_M\cdot\vda_N
=\int\d\vr \vda_M\Rht\vda_N$ subject to
$\int\d\vr \vda_M\cdot\vda_N=\delta_{MN}$.
Using \eqref{eqn:dDefKS} the two integrals become
\begin{align}
\int\d\vr \vda_M^*\cdot\vda_N=&\sum_{jk} p_{Mj}^*p_{Nk} N_{jk},
\\
\int\d\vr \vda_M^*\Rht\vda_N=&\sum_{jk} p_{Mj}^*p_{Nk} R_{jk}.
\end{align}
where
\begin{align}
N_{jk}=&\int\d\vr
[(\Da-\Ra)\psi_j^*][(\Da-\Ra)\psi_k],
\\
R_{jk}=&\int\d\vr [(\Dm-\Rm)\psi_j^*]\Rh_{\mu\nu}
[(\Dn-\Rn)\psi_k].
\label{eqn:RhDefKS}
\end{align}
Minimising with respect to $p_{Mj}$,
the eigen-equation \eqref{eqn:QCMEigd} thus becomes
\begin{align}
\Omegat_N^2N_{jk}p_{Nk}=&R_{jk}p_{Nk}
\end{align}
subject to the orthogonality condition
$\sum_{jk} N_{jk}p_{Mj}^*p_{Nk}=\delta_{NM}$.
Following the details of \Aref{app:KS} we find
\begin{align}
R_{jk}=&\int\d\vr [\Dfm\psi_j^*]V_{,\mu\nu}[\Dfn\psi_k]
\nonumber\\&
+ \frac14 \int\d\vr [\Dfa\Dfm\Dfn\psi_j^*]
[\Dfa\Dfm\Dfn\psi_k]
\nonumber\\&
+ 3\int\d\vr [\Dfa\Dfm\psi_j^*]
t_{\mu\nu}[\Dfa\Dfn\psi_k]
\end{align}
where $\Dfa\equiv\Da-\Ra$.

\subsection{Matrix form of the scalar approximation}

Since all terms involve repeated use of the operator $(\Da-\Ra)$
we may simplify things somewhat by adopting a matrix notation
to represent this operator.
For a finite basis set of size $\NBas$ we
define the following $\NBas\times\NBas$ matrices: $\mD_{\alpha}$,
$\mT_{\mu\nu}$ and $\mV_{\mu\nu}$ with elements
\begin{align}
D_{\alpha jk}=&\int\d\vr \psi_j^*(\Da-\Ra)\psi_k,
\\
T_{\mu\nu jk}=&\int\d\vr \psi_j^*t_{\mu\nu}\psi_k,
\\
V_{\mu\nu jk}=&\int\d\vr \psi_j^*V^{\KS}_{,\mu\nu}\psi_k.
\end{align}
We can also integrate by parts to obtain
$V_{\mu\nu jk}=\int\d\vr V^{\KS}[\Dm\Dn\psi_j^*\psi_k]$.
Noting that the orbitals $\psi_j$ form an orthonormal set
we see that $(\Da-\Ra)\psi_k=\sum D_{\alpha jk} \psi_j$
with similar relationships for the others.

This allows us to write the matrix eigen-equation
\begin{align}
\Omegat_N^2\mat{N}\mat{P}_N=&\mat{R}\mat{P}_N
\label{eqn:QCMEigKS}
\end{align}
where $\mat{P}_N$ is an $\NBas\times 1$ matrix with elements $p_{Nj}$.
Here
\begin{align}
\mat{N}=&\mD_{\alpha}^{\dag}\mD_{\alpha},
\label{eqn:NKS}
\\
\mat{R}=&\mD_{\mu}^{\dag}\mV_{\mu\nu}\mD_{\nu}
+ \mD_{\mu}^{\dag}\mD_{\alpha}^{\dag}
\lbr 3\mT_{\mu\nu}+\frac14 \mD_{\nu}^{\dag}\mD_{\mu} \rbr
\mD_{\alpha}\mD_{\nu}.
\label{eqn:RKS}
%+ \frac14 \mD_{\alpha}^{\dag}\mD_{\mu}^{\dag}\mD_{\nu}^{\dag}
%\mD_{\nu}\mD_{\mu}\mD_{\alpha}
%\nonumber\\&
%+ 3\mD_{\alpha}^{\dag}\mD_{\mu}^{\dag}\mT_{\mu\nu}\mD_{\nu}\mD_{\alpha}
\end{align}
Orthogonality is given by
$\mat{P}_N^{\dag}\mat{N}\mat{P}_M=\delta_{NM}$.
We note that all tensor components are bundled into $\mat{N}$ and $\mat{R}$
via summation over $\alpha$, $\mu$ and $\nu$.
These equations are only true in the strict limit of 
an infinite number of orbitals but will converge to the
true approximate solution in typical systems.

\subsection{One-electron like asymptotic behaviour}
\label{ssec:OneElAs}

In the asymptotic $|\vr|\to\infty$ region
of a localised system, the
highest occupied molecular orbital (HOMO) $\psi_H$
with occupation number $f_H$
dominates [at least in the non-degenerate case or for degenerate
system with spherical symmetry where we set $f_H=(2l_H+1)$
and $\psi_H(\vr)=R_{n_H}(r)$, the radial component of
$\psi_{n_Hl_Hm}(\vr)=R_{n_H}(r)Y_{l_Hm}(\Omega_r)$]. Thus,
in the limit $\vr\to\infty$ we can set:
\begin{align}
\n\aeq& f_H|\psi_H|^2,
&
\Rv\aeq& \frac{[\nabla \psi_{H}]}{\psi_H},
\\
t_{\mu\nu}\aeq&-\half\Dm\Rn,
&
S\aeq& 2(V^{\KS}-\epsilon_H).
\end{align}
All equations reduce to the one-electron form in these limits,
with the single occupied orbital given by the HOMO.

We can check how close a region of a many-electron system is
to a one-electron-like system by considering the
difference of $S$ from its asymptotic form. Here we
set
\begin{subequations}
\begin{align}
\tilde{S}(\vr)=&S(\vr)-2(V^{\KS}(\vr)-\epsilon_H)
\\
=&C^0(\vr)+2(\epsilon_H-\epsilont(\vr))
\end{align}\label{eqn:Stilde}%
\end{subequations}
where
\begin{align}
C^0(\vr)=&\sum_i \frac{f_i|\nabla\psi_i(\vr)|^2}{\n(\vr)}-|\Rv(\vr)|^2
\\
\epsilont(\vr)=&
\sum_i \frac{f_i|\psi_i(\vr)^2|}{\n(\vr)}\epsilon_i
\end{align}
and where $C^0$ is related to the
Fermi-hole curvature\cite{Dobson1991,*Dobson1993} of the KS system.
In the limit $|\vr|\to\infty$ it is clear that $C^0(\vr)=0$ and
$\epsilont(\vr)=\epsilon_H$ (true everywhere for one-electron systems).

This outer region is crucial for van der Waals physics in many systems.
Typically only transitions within a small frequency range,
especially that to the lowest unoccupied molecular orbital (LUMO),
will dominate in this region. In such transitions
$\int\d\vr\psi_k^*\tilde{S}\psi_k$ will be small.
The asymptotic `exactness' in this limit suggest strongly that
the scalar approximation to QCM is a appropriate for calculating vdW forces.

Even a well chosen but simple approximation,
using a limited number of unoccupied orbitals and quasi-orbitals,
may provide quite accurate estimations of the vdW physics.
The ability of the QCM (and the scalar approcimation)
to make collective modes for the transition physics may
aid in convergence compared to full dRPA calculations.

\subsection{Transitions and energy calculations in the scalar approximation}

Once we have obtained solutions of \eqref{eqn:QCMEigKS}
we can use $p_{Nj}$ to evaluate $\DN$ through \eqref{eqn:dNKS}.
Using $\Omegat_N$ also we can then evaluate $\chih_0$ and 
properties which depend on it such as the correlation energy.

We can also use the asymptotic form of $S(\vr)$
to further investigate $\DN$ and through it correlations.
Combining \eqref{eqn:dNKS} and \eqref{eqn:Stilde} lets us write
\begin{align}
\DN=&\Psi^0\sum_jp_{Nj}
\lbrs 2(V^{\KS}-\epsilon_H)-\nabla^2+\tilde{S}\rbrs\psi_j
\nonumber\\
=&\Psi^0\sum_jp_{Nj} \lbr 2\Omega_{jH}+\tilde{S}\rbr\psi_j
\label{eqn:dNSA}
\end{align}
where the $\Omega_{jH}=\epsilon_j-\epsilon_H$ comes from applying
the Schr\"odinger Hamiltonian $-\half\nabla^2+V^{\KS}$ to $\psi_j$.
Here $\tilde{S}$ covers the deviation
of many-electron systems from their one-electron equivalents.

Let us expand \eqref{eqn:dNSA} as follows:
\begin{align}
\DN=&\Psi^0\sum_j d_{Nj}\psi_j.
\label{eqn:dNKSD}
\end{align}
We define matrices $\mat{O}$ with $O_{jk}=\delta_{jk}\Omega_{kH}$
and $\tilde{\mat{S}}$ with
$\tilde{S}_{jk}=\int\d\vr \psi_j^*\tilde{S}\psi_k$.
Thus $d_{Nj}=[\mat{D}_N]_j$
where $\mat{D}_N=\lbr 2\mat{O} + \tilde{\mat{S}} \rbr\mat{P}_N$
and $\mat{P}_N$ is a solution of \eqref{eqn:QCMEigKS}.
We can use this expansion to calculate the Coulomb projections
$W_{NM}$ defined in \eqref{eqn:WNM}.
Using \eqref{eqn:dNKSD} we find
\begin{align}
W_{NM}=&\mat{D}_M^{\dag}\mat{W}^{(\psi)}\mat{D}_N
\nonumber\\
=&\mat{P}_M^{\dag}(2\mat{O}+\tilde{\mat{S}})\mat{W}^{(\psi)}
(2\mat{O}+\tilde{\mat{S}})\mat{P}_N
\end{align}
where 
\begin{align}
[\mat{W}^{(\psi)}]_{jk}=&\int\d\vr\d\vr
\frac{\Psi^0(\vr)\psi_j^*(\vr)\Psi^0(\vrp)\psi_k(\vrp)}%
{|\vr-\vrp|}.
\end{align}
We can use $W_{NM}$ to calculate the correlation energy
through equation \eqref{eqn:Ec}.

In molecular orbital language it is clear that the calculation
of $\mat{W}^{(\psi)}$ is the only step involving four-center integrals
required by a correlation energy calculation.
However, in contrast to the four orbital (two occupied, two unoccupied)
terms $(ia|jb)$ required by a full KS calculation it
requires two uncontracted orbital indices (occupied or unuccopied) only,
provided we can efficiently and accurately project $\Psi^0=\sqrt{\n}$
onto a fitting basis set. For large molecules with many occupied
orbitals this is a substantial saving.

\section{1D Harmonic oscillator}
\label{sec:Harmonic}

To illustrate the approaches discussed here, let us investigate the
case of a one-dimensional (1D) many-electron Harmonic oscillator
without interactions in the groundstate.
The system is thus defined as having $N_e$ electrons
in the groundstate and $V^{\KS}(x)=x^2/2$.
Here the KS orbital wavefunctions and energies take the form
\begin{align}
\psi_j(x)=&\kappa_jH_j(x)e^{-x^2/2},
&
\epsilon_j=&\frac{2j+1}{2}
\end{align}
where $H_j(x)$ are Hermite polynomials with $j\geq 0$
and $\kappa_j=[\sqrt{\pi}2^jj!]^{-1/2}$.
Since the orbitals are filled up to $j\leq N_e-1$ the density is
\begin{align}
\n(x)=&\sum_{j=0}^{N_e-1} \frac{H_j(x)^2e^{-x^2}}{\sqrt{\pi}2^jj!}
\end{align}
and it is clear from the expansion of $H_{N_e-1}$
that $\lim_{x\to\infty} \n(x)\sim x^{2(N_e-1)}e^{-x^2}$ and
$\lim_{x\to\infty} \Rs(x)\sim -x$. Thus the
system is well-bounded and the KS-orbital approach
should work well.

In the one-electron case $\Rs=-x$, $t=\half$ and $V_{,xx}=1$.
and we can solve everything analytically.
The matrices required to form $\mat{N}$ and $\mat{R}$
in equations \eqref{eqn:NKS} and \eqref{eqn:RKS}
have elements $T_{jk}=\half \delta_{jk}$,
$V_{jk}=\delta_{jk}$ and
\begin{align*}
D_{jk}=&\kappa_j\kappa_k\int\d x e^{-x^2/2} H_j(x)[(\Dp{x}+x)H_k(x)e^{-x^2/2}]
\\
=&2k\kappa_j\kappa_k\int\d x e^{-x^2} H_j(x)H_{k-1}(x)
\\
%=&2(j+1)\frac{\kappa_{j+1}}{\kappa_j}\delta_{j(k-1)}
%\\
=&\sqrt{2(j+1)}\delta_{j(k-1)}.
\end{align*}
Here $[\mD^{\dag}\mD]_{jk}=2j\delta_{jk}$,
$[\mD^{\dag 2}\mD^2]_{jk}=4j(j-1)\delta_{jk}$
and $[\mD^{\dag 3}\mD^3]_{jk}=8j(j-1)(j-2)\delta_{jk}$.

Thus $N_{jk}=2j\delta_{jk}$ and
$R_{jk}=2j\delta_{jk} + 6j(j-1)\delta_{jk} + 2j(j-1)(j-2)\delta_{jk}
=2j^3\delta_{jk}$.
We can ignore the $j=0$ solution as it will not contribute to $\chih_0$.
We thus choose solutions with $N_j>0$ where $p_{N_jj}=\delta_{N_jj}/\sqrt{2j}$.
Here $\Omega_{N_j}=\frac{2j+1}{2}-\half=j$
as expected and the transition density mode
is $\DNj(x)=-\psi_0(x)[(\Dx^2+1-x^2)\psi_j(x)]/\sqrt{2j}
=\sqrt{2\Omega_{N_j}}\psi_0(x)\psi_j(x)$.

In the many-electron case we can solve the problem
semi-analytically for much of it, requiring
numerics only for terms involving $1/\n$ and subsequent
diagonalisations. Errors occur due to truncation of
the basis set but integrals can be obtained
with near exact accuracy with Gauss-Hermite quadrature.
It should be noted that the two-electron system is not
predicted exactly by the QCM as there is no spin degeneracy.

The rapid convergence of the method is demonstrated in
Table~\ref{tab:Converge} where we show the convergence
of the fourth transition frequency $\Omega_4$ for the
two-electron system. With as few as 15 states the error
is already under one part in ten thousand.
\begin{table}[hb]
\begin{tabular*}{0.95\columnwidth}{@{\extracolsep{\fill}}|c|rrrrrr|}
\hline
$\NBas$ & 5 & 10 & 15 & 20 & 50 & $\infty$
\\\hline
$\Omega_{4}$ & 4.1105 & 3.8797 & 3.8805 & 3.8801 & 3.8802 & 3.8802
\\
$\log_{10}|\rm{Err}|$ & -1.22 & -3.95 & -4.05 & -5.06 & -7.49 & $-\infty$
\\
\hline
\end{tabular*}
\caption{Convergence of the fourth transition frequency
for the two electron system. Here $\rm{Err}=\Omega_4^{\NBas}/\Omega_4^{\infty}-1$.
\label{tab:Converge}}
\end{table}

The KS transition frequencies of an $N_e$ electron system
have energies $\Omega_{ai}=\frac{2a+1}{2}-\frac{2i+1}{2}=J$
and each frequency $J$ has multiple contributing transitions.
Here $\min(J,N_e)$ modes have transition frequency $J$
and the density transitions are  proportional to
$\psi_i(x)\psi_{i+J}(x)$ when $i\leq N_e$ and $i+J> N_e$.

We present the QCM transition frequencies in
Table~\ref{tab:Omega} for systems with up to 20 electrons.
The QCM has \emph{single-valued} frequencies distributed
approximately with the integers. Each QCM
transition density is therefore composed of multiple
KS-like transitions which, by the sum rules \eqref{eqn:SumRules}
must have their weights $|K_{aiN}|^2$ dominated by transitions
with $\Omega_{ai}\aeq\Omega_N$.
In Table \ref{tab:HaiN} we show some weights $|K_{aiN}|^2$ for the
five electron system. It is clear that the $N$th QCM transition
puts most weight on the
HOMO-HOMO+$N$ [$5-(5+N)$] KS transition as one would hope.

\begin{table}[th]
\begin{tabular*}{0.95\columnwidth}{@{\extracolsep{\fill}}|rrrrrr|}
\hline
\multicolumn{6}{|c|}{Transition frequencies $\Omega_N$}
\\
$N$ & $N_e=1$ & $N_e=2$ & $N_e=5$ & $N_e=10$ & $N_e=20$
\\\hline
1  & 1.0000 & 1.0000 & 1.0000 & 1.0000 & 1.0000 \\
2 & 2.0000 & 2.0000 & 2.0000 & 2.0000 & 2.0000 \\
3 & 3.0000 & 3.0000 & 3.0000 & 3.0000 & 3.0000 \\
4 & 4.0000 & 3.8802 & 3.9531 & 3.9859 & 3.9963 \\
5 & 5.0000 & 4.8680 & 4.8162 & 4.9225 & 4.9772 \\
6 & 6.0000 & 5.7877 & 5.6869 & 5.7886 & 5.9216 \\
7 & 7.0000 & 6.7689 & 6.6309 & 6.6381 & 6.8128 \\
8 & 8.0000 & 7.7154 & 7.5079 & 7.5381 & 7.6683 \\
10 & 10.0000 &  9.6579 &  9.3578 &  9.3030 &  9.4403 \\
20 & 20.0000 & 19.4882 & 18.6483 & 18.2017 & 18.1953 \\
\hline
\end{tabular*}
\caption{First eight, 10th and 20th
distinct eigenfrequencies of the KS system
and QCM system with different numbers of electrons $N_e$.
The KS transitions are all integer
and degenerate with $\min(\Omega,N_e)$
transitions for a given integer frequency.\label{tab:Omega}}
\end{table}

\begin{table}[bh]
\begin{tabular*}{0.95\columnwidth}%
{@{\extracolsep{\fill}}|rr|rrrrrr|}
\hline
$N$ & $\Omega_N$ &
($i$,$a$) & $|K_{aiN}^2|$ & ($i$,$a$) & $|K_{aiN}^2|$ &
($i$,$a$) & $|K_{aiN}^2|$ 
\\\hline
1 & 1.0000 & (5,6) & 100.0
&&&&\\
2 & 2.0000 & (5,7) & 60.0 & (4,6) & 40.0
&&\\
3 & 3.0000 & (5,8) & 53.8 & (4,7) & 30.8 & (3,6) & 15.4
\\
4 & 3.9531 & (5,9) &  54.4 & (4,8) &  27.1 & (3,7) &  11.5 
\\&& (2,6) &   3.8 & (4,6) &   1.9 & (5,7) &   1.2
\\
5 & 4.8162 & (5,10) &  52.8 & (4,9) &  23.9 & (3,8) &   9.0
\\&& (3,6) &   6.2 & (5,8) & 4.2 & (2,7) &   2.6 
\\
7 & 6.6309 & (5,12) &  49.8 & (4,11) &  20.0 & (5,10) &   6.8
\\&& (3,10) &   6.7 & (3,8) &  5.7 & (2,7) &   5.2
\\&& (1,6) &   1.9 & (2,9) &   1.7
&&\\
10 & 9.3578 & (5,15) &  45.0 & (4,14) &  16.5 & (5,13) &  14.2
\\&& (3,13) &   5.0 & (3,11) &   4.0 & (2,10) &   3.2
\\&& (2,8) &   3.1 & (1,7) &   2.3 & (5,17) &   1.1 
\\&& (2,12) &   1.1
&&&&\\\hline
\end{tabular*}
\caption{Weights $|K_{aiN}|^2$ in $\%$ with $>1\%$ contribution
for the five electron system. The KS transition frequencies
are $\Omega_{ai}=a-i$. Tabulated weights may not sum to $100\%$
due to the absence of $<1\%$ contributions.\label{tab:HaiN}}
\end{table}

It is remarkable that the lowest three frequencies are
exact to four decimal places for $N_e>1$,
particularly as the second smallest is comprised of two transitions
and the third of three in all but the two-electron case.
By the sum rules \eqref{eqn:SumRules} this means that
each mode must be comprised solely
of transitions with the given frequency, and as a consequence,
both $\Omega_N$ and $\DN$ are predicted exactly for
the HOMO-LUMO transition.
That such a relationship holds for as many
as 20 electrons demonstrates a strength of the physics
and a resilience of the approximations in the QCM approach.

\section{Conclusion}

In this paper we have reformulated, simplified and investigated the
QCM formalism of \rcites{Tokatly2007,Tao2009,Gao2010,Gould2011},
and provided a direct proof of its exactness
in one-electron systems.

Firstly, in \sref{sec:QCMForm} we provide a more comprehensive derivation of,
and investigation into the compact form of the QCM
equations, especially with regards to the kinetic stress tensor
defined by \eqref{eqn:T0mn}. The simplified QCM equation
is described in \eqref{eqn:QCMF} and surrounding
work. We then discuss the QCM response function
[\eqref{eqn:Chi0} in \sref{sec:Response}]
and use it to derive a simple expression for the
QCM-dRPA correlation energy in \eqref{eqn:EcSum}.

The orthonormal form described in \sref{sec:Ortho}
is vital for the description and calculation
of localised systems, where the standard $\vu$-based formulation
does not behave well. This reformulation can be written
in a straightforward manner [see especially \eqref{eqn:Rhd}]
and is required for
calculation of atomic, and molecular systems using standard
basis set approaches involving GTO and STOs.
It also provides a relatively simple way of proving (in \sref{sec:OneEl}),
direct from the Schr\"odinger equation, that the QCM
is exact for one-electron systems.

The scalar approximation of \sref{sec:ScalarQCM} is then
derived from the one-electron case.
Is is exact in one-electron systems
and should accurately predict the physics of asymptotic
regions where the behaviour is essentially one-electron like.
Scalar QCM is then conveniently expressed in \eqref{eqn:QCMEigphi},
which provides a numerically simpler approximation to the full QCM,
with some sacrifice of exact properties of the full theory in general
systems. The scalar approximation could be used as a
faster alternative to full QCM itself or
as a doorway to further and cruder approximations.
In particular it opens up analysis of transition
frequencies and densities in a simple manner by
relating information from the KS orbitals
to the collective modes in the QCM.

Finally we test the approaches developed in this manuscript
on a simple one-dimensional many-electron in \sref{sec:Harmonic}.
Results for this test system are generally excellent,
demonstrating again the ability of the QCM to include the
important physics of a many-electron system through the
use of the displacement $\vu$ only (or the equivalent $\vd$).

Both reformulations presented here will aid in the development
of a robust, basis set based, QCM approach for atomic and
molecular systems. The difficulty in such approaches
is the presence of non-constant denominators in
$t_{\mu\nu}$, $\Ra$ and $S$. We believe, however, that
a tractable means of dealing with these
should be possible by using Gauss-Hermite quadrature with GTOs.

It should also be possible to further develop vdW
functionals that use a further approximation to the QCM based on
cleverly chosen orbital-like basis functions, possibly incorporating
some unoccupied KS orbitals. Such functionals could
incorporate ideas from \rcites{Grimme2006,Tkatchenko2009,Vydrov2009}
to be made even more efficient than the functional described
here or in \rcite{Gould2011}, without sacrificing the vital
high-frequency, non-additive, and long-ranged physics that make
the QCM attractive.

\acknowledgments

We would like to thank Janos Angyan, Andreas Savin
and Giovanni Vignale for much fruitful discussion.
G. J. thanks John Dobson for his hospitality and support during a
sabbatical stay in Brisbane.
J.F.D. and T.G. were supported by ARC Discovery Grant DP1096240.
I. V. T. was supported by the Spanish MICINN,
Grant No. FIS2010-21282-C02-01, and
``Grupos Consolidados UPV/EHU del Gobierno Vasco,'' Project No. IT-319-07.

\appendix
\section{Kinetic stress tensor}
\label{app:Kin}

There is some debate over the appropriate form of 
the kinetic stress tensor $\ten{T}^{\Kin}$.
We have chosen the form given in 
\eqref{eqn:TmnGen} and \eqref{eqn:T0mn}, whereas the original
QCM papers used the form given in \eqref{eqn:TTokGen}.
One requirement is that, for Kohn-Sham systems,
$\tT$ obeys the force balance condition
$\vnabla\cdot\tT=-\n\nabla V^{\KS}$ which is true both for
the form \eqref{eqn:T0mn} used in this manuscript and for
equation 17 of \rcite{Tao2009} [leading to our equation 
\eqref{eqn:T0Tok} for a Kohn-Sham groundstate].

There are various ways to arrive at the possible forms of
$\ten{T}^{\Kin}(\vr)$
(see e.g. \rcites{Schroedinger1927,PauliHandbook1933,%
FeynmanBScThesis1939,Martin1959,Puff1968}).
For completeness, we show one way to arrive at the form
\eqref{eqn:TmnGen}, \eqref{eqn:T0mn} chosen for the present work. 
This form helps to make the QCM equations particularly simple.
We base our form on the Wigner transform of the ``classical''
kinetic stress tensor. In a ``classical'' fluid with no 
current present, $\ten{T}^{\Kin}$ is defined by
\begin{align}
T^{\Kin}_{\mu\nu}(\vr)=&\int\d\vp p_{\mu}p_{\nu} f(\vr,\vp).
\label{eqn:T0mnClass}
\end{align}
%%%inserted 21.01.2012 by JFD
Here $f$ is the classical one-body distribution function
- i.e. the phase space probability 
density for finding a particle at position $\vr$ with momentum $\vp$.
To model a quantum system we can replace
$f(\vr,\vp)$ by the Wigner form
\begin{align}
f(\vr,\vp)=&\int\frac{\d\vx}{(2\pi)^3}
\rho\lbr \vr+\frac{\vx}{2}, \vr-\frac{\vx}{2} \rbr
e^{i\vp\cdot\vx}
\end{align}
where $\rho (\vr,\vrp)$ is the quantal one-body density matrix.
Using the short-hand $\rho=\rho(\vr+\vx/2,\vr-\vx/2)$
we thus find
\begin{align}
%%% \T replaced by T on purpose 
T^{\Kin}_{\mu\nu}(\vr)=&\int\d\vp
\int\frac{\d\vx}{(2\pi)^3}\rho
p_{\mu}p_{\nu} e^{i\vp\cdot\vx}
\nonumber\\
=&\int\d\vp
\int\frac{\d\vx}{(2\pi)^3}\rho
[-\Dp{x_{\mu}}\Dp{x_{\nu}} e^{i\vp\cdot\vx}]
\nonumber\\
=&\int\frac{\d\vx}{(2\pi)^3}
[-\Dp{x_{\mu}}\Dp{x_{\nu}} \rho ] \int\d\vp e^{i\vp\cdot\vx}
\nonumber\\
=&[-\Dp{x_{\mu}}\Dp{x_{\nu}} \rho ]_{\vx=\vec{0}},
\end{align}
and so, by the chain rule,
\begin{align}
T^{\Kin}_{\mu \nu}(\vr)&= -\Dp{x_{\mu}}\Dp{x_{\nu}}
\rho(\vr+\vx/2,\vr-\vx/2)|_{\vx=\vec{0}}
\nonumber\\
&=\frac14 [(\Dm\Dn' + \Dm'\Dn - \Dm\Dn - \Dm'\Dn')\rho(\vr,\vrp)]_{\vr=\vrp}
\label{eqn:TasDeriv}
\end{align}
with $\Dm\equiv\Dp{r_{\mu}}$ and $\Dm'\equiv\Dp{r_{\mu}'}$.
%%%added w21.01.2012 by JFD
Furthermore
\begin{align}
&\frac{1}{4}\left[ \left( \Dm \Dn + \Dm \Dn \pr
+\Dm\pr\Dn + \Dm\pr \Dn\pr \right) \rho (\vr,\vrp)\right]_{\vr=\vrp}
\nonumber\\&
=\frac{1}{4}\Dm \Dn \rho (\vr,\vr)
=\frac{1}{4}\Dm \Dn n(\vr)
\label{eqn:NMuNu}
\end{align}%
Combining \eqref{eqn:TasDeriv} and \eqref{eqn:NMuNu} we find
\begin{align}
T^{\Kin}_{\mu \nu}(\vr) =& \frac{1}{2}\left[ (\Dm \Dn\pr + \Dm\pr \Dn  )
\rho(\vr,\vrp)\right]_{\vr=\vrp}
\nonumber\\&
-\frac{1}{4}\Dm \Dn n(\vr) .
\end{align}
for a state with zero current, in agreement with \eqref{eqn:TmnGen}.

Since $\rho(\vr,\vrp)=\sum_i f_i\psi_i^*(\vr)\psi_i(\vrp)$ in
a groundstate KS system we find the stress tensor $\tT$ for
this situation to be
\begin{align}
\T_{\mu\nu}(\vr)=&
\frac12\sum_i f_i(\Dm\Dn' + \Dm'\Dn)\psi_i^*(\vr)\psi_i(\vrp)|_{\vr=\vrp}
\nonumber\\&
- \frac14[\Dm\Dn \n(\vr)]
\nonumber \\
%-\frac14\sum_i f_i(\Dm\Dn + \Dm'\Dn')\psi_i^*(\vr)\psi_i(\vrp)|_{\vr=\vrp}
%\nonumber\\
%=&\half\Re\sum_i f_i\lbr[\Dm\psi_i^*][\Dn\psi_i]-\psi_i^*[\Dm\Dn\psi_i]\rbr
\nonumber\\
=&\Re\sum_i f_i[\Dm\psi_i^*][\Dn\psi_i] - \frac14[\Dm\Dn \n]
\end{align}
Thus \eqref{eqn:TmnGen} is exactly the Wigner version of the
``classical'' kinetic stress tensor \eqref{eqn:T0mnClass}.

Using \eqref{eqn:T0mnA} %rather than \eqref{eqn:T0mn}
we demonstrate that $\Da\T_{\alpha\mu}=-\n V^{\KS}_{,\mu}$.
Here we note that
\begin{align}
\Da\T_{\alpha\mu}=\half \sum_j f_j t_{j\mu}
\end{align}
where
\begin{align*}
t_{j\mu}=&\Re\big\{
[\nabla^2\psi_j^*][\Dm\psi_j] + [\Da\psi_j^*][\Da\Dm\psi_j]
\nonumber\\&
- [\Da\psi_j^*][\Da\Dm\psi_j] - \psi_j^*[\Dm\nabla^2\psi_j]
\big\}
\\
=&\Re\big\{
[\nabla^2\psi_j^*][\Dm\psi_j]
- \psi_j^*[\Dm\nabla^2\psi_j]
\big\}
\\
%=&2\Re\big\{\psi^*_{j,\mu}(V^{\KS}-\epsilon_j)\psi_j
%-\psi_j^*(V^{\KS}-\epsilon_j)\psi_{j,\mu}
%\nonumber\\&
%-\psi_j^*V^{\KS}_{,\mu}\psi_j
%\big\}
%\\
=&-2V^{\KS}_{,\mu}|\psi_j|^2
\end{align*}
and we have used the Schr\"odinger equation
$\nabla^2\psi_j=2(V^{\KS}-\epsilon_j)\psi_j$
to derive the final expression. Finally
$\Da\T_{\mu\alpha}=-\sum_j f_j V^{\KS}_{,\mu}|\psi_j|^2=-\n V^{\KS}_{,\mu}$
and the proof is complete.

Comparing this with the form $\tTb$ \eqref{eqn:T0Tok} used in the
original QCM formulation, we find
$\Da \Tb_{\mu\alpha}=\Da \T_{\mu\alpha}$ since
$\Da [\Dm\Da\n]=[\Dm\nabla^2\n]
=\Da \delta_{\mu\alpha}[\nabla^2\n]$. Since these are precisely
the terms that differ between \eqref{eqn:T0mn} and \eqref{eqn:T0Tok}
it follows that the gradients must be identical.

\section{Compact form of the QCM equations}
\label{app:Compact}

In an earlier work\cite{Gould2011} we state without proof that
equation \eqref{eqn:QCMu} defined via \eqref{eqn:RHat} here
(equations 2-5 of \rcite{Gould2011})
are equivalent to equations 14-16 of \rcite{Tao2009}.
The only non-notational difference is in the kinetic force term
$\vF^{1\Kin}=-\Kht\vu$ defined here via \eqref{eqn:KHat},
in equation 14 of \rcite{Tao2009}, and in
equation 53 of \rcite{Gao2010} (abbreviated
as G53) where it is derived from their equation C8 (abbreviated as GC8).

We demonstrate that the two forms are equivalent by working from GC8.
While the same result can be obtained directly from G53 the
derivation is less clear and less succinct. Specifically
we must show that
\begin{align}
\vF^{1\Kin}_{\mu}\equiv&\frac{\delta \FT_2[\vu]}{\delta u_{\mu}}=
-\Kh_{\mu\nu}u_{\nu}
\label{eqn:dT2u}
\end{align}
where $\Kh_{\mu\nu}$ is defined in \eqref{eqn:KHat},
since the remaining terms in \eqref{eqn:QCMu} follow directly
from equations 14-16 of \rcite{Tao2009}.

Following Gao~et~al\cite{Gao2010} we write GC8 as
$\FT_2[\vu]=\int\d\vr \FI$ where
\begin{align}
\FI=&\Big\{
\FK_{\mu\nu}(4U_{\mu\alpha}U_{\nu\alpha}-u_{\alpha,\mu}u_{\alpha,\nu})
+ \frac{\n}{8}U_{\alpha\alpha,\mu}U_{\nu\nu,\mu}
\nonumber\\&
+ \frac{\n_{,\nu}}{2}U_{\mu\nu}\Dm U_{\alpha\alpha}
+ \frac{\n_{,\nu}}{4}u_{\alpha,\mu}\Dn u_{\mu,\alpha}
\Big\},
\label{eqn:GC8}
\end{align}
with
$U_{\mu\nu}=\half( u_{\mu,\nu} + u_{\nu,\mu} )$, $U_{\alpha\alpha}=u_{\alpha,\alpha}$,
and $\FK_{\mu\nu}=\half\Re \sum_i \allowbreak f_i[\Dm\psi_i^*][\Dn\psi_i]$.

We can use integration by parts to remove derivatives of $\n$
from the integrand $\FI$. As such
\begin{align}
\FT_2[\vu]=&\half\int\d\vr
\lbr 2\FK_{\mu\nu}\FY_{\mu\nu} - \frac{\n}{4} \FZ \rbr
\nonumber\\
=&\half\int\d\vr \lbrs \T_{\mu\nu}\FY_{\mu\nu}
+ \frac{\n}{4} (\FY_{\mu\nu,\mu\nu}-\FZ) \rbrs
\label{eqn:T2P}
\end{align}
where we define $\T_{\mu\nu}=2\FK_{\mu\nu}-\frac14 \n_{,\mu\nu}$ to be equal
to \eqref{eqn:T0mn} rather than the form appearing
in \rcites{Tao2009,Gao2010}. The undefined terms in the
integrand of \eqref{eqn:T2P} take the form
\begin{align}
\FY_{\mu\nu}=&u_{\mu,\alpha}u_{\nu,\alpha} + u_{\alpha,\mu}u_{\nu,\alpha}
+ u_{\mu,\alpha}u_{\alpha,\nu},
\\
\FZ=& u_{\mu,\mu\alpha} u_{\nu,\nu\alpha}
+ 2u_{\mu,\nu\alpha} u_{\nu,\mu\alpha}
\nonumber\\&
+ u_{\mu,\nu\alpha\alpha} u_{\nu,\mu}
+ u_{\mu,\nu} u_{\nu,\mu\alpha\alpha}
\nonumber\\&
+ u_{\mu,\alpha\alpha} u_{\nu,\mu\nu}
+ u_{\mu,\nu\mu} u_{\nu,\alpha\alpha}
\nonumber\\&
+ 2 u_{\mu,\mu\nu\alpha} u_{\nu,\alpha}
+ 2 u_{\mu,\alpha} u_{\nu,\mu\nu\alpha}
\end{align}
where we have expanded $U_{\mu\nu}$ 
and used the product rule on all derivatives
to arrive at these forms.
The following identities are also used in the derivation of $\FZ$:
\begin{align*}
U_{\alpha\alpha,\mu}U_{\nu\nu,\mu}
=&u_{\mu,\mu\alpha}u_{\nu,\nu\alpha}
\\
2\Dn u_{\mu,\nu}\Dm u_{\alpha,\alpha}
=&\Da u_{\mu,\alpha}\Dm u_{\nu,\nu} + \Da u_{\nu,\alpha}\Dn u_{\mu,\mu},
\\
2\Dn u_{\nu,\mu}\Dm u_{\alpha,\alpha}
=&\Dn u_{\nu,\alpha}\Da u_{\mu,\mu} + \Dm u_{\mu,\alpha}\Da u_{\nu,\nu},
\\
2\Dn u_{\alpha,\mu} \Dn u_{\mu,\alpha}
=&\Da u_{\nu,\mu} \Da u_{\mu,\nu} + \Da u_{\mu,\nu} \Da u_{\nu,\mu}
\end{align*}
which follow from exchange of indices under summation
(eg. $A_{\nu\mu}B_{\mu\nu}\equiv A_{\mu\nu}B_{\nu\mu}$).

We can expand the terms of $\FY_{\mu\nu,\mu\nu}$ as follows:
\begin{align*}
\Dm\Dn u_{\mu,\alpha}u_{\nu,\alpha}
=&u_{\mu,\mu\alpha}u_{\nu,\nu\alpha} + u_{\mu,\nu\alpha}u_{\nu,\mu\alpha}
\nonumber\\&
+ u_{\mu,\mu\nu\alpha}u_{\nu,\alpha} + u_{\mu,\alpha}u_{\nu,\mu\nu\alpha},
\\
\Dm\Dn u_{\mu,\alpha}u_{\alpha,\nu}
=& u_{\mu,\mu\nu}u_{\nu,\alpha\alpha} + u_{\mu,\nu\alpha}u_{\nu,\mu\alpha}
\nonumber\\&
+ u_{\mu,\mu\nu\alpha}u_{\nu,\alpha} + u_{\mu,\nu}u_{\nu,\mu\alpha\alpha},
\\
\Dm\Dn u_{\nu,\alpha}u_{\alpha,\mu}
=& u_{\mu,\alpha\alpha}u_{\nu,\mu\nu} + u_{\mu,\nu\alpha}u_{\nu,\mu\alpha}
\nonumber\\&
+ u_{\mu,\alpha}u_{\nu,\mu\nu\alpha} + u_{\mu,\nu\alpha\alpha}u_{\nu,\mu}
\end{align*}
where we again exchange indices where appropriate.
This leads to the following result
\begin{align}
\FY_{\mu\nu,\mu\nu}=&u_{\mu,\mu\alpha}u_{\nu,\nu\alpha} + 3u_{\mu,\nu\alpha}u_{\nu,\mu\alpha}
\nonumber\\&
+ u_{\mu,\alpha\alpha}u_{\nu,\mu\nu} + u_{\mu,\mu\nu}u_{\nu,\alpha\alpha}
\nonumber\\&
+ u_{\mu,\nu\alpha\alpha}u_{\nu,\mu} + u_{\mu,\nu}u_{\nu,\mu\alpha\alpha}
\nonumber\\&
+ 2u_{\mu,\alpha}u_{\nu,\mu\nu\alpha} + 2u_{\mu,\mu\nu\alpha}u_{\nu,\alpha}
\\
=&\FZ + u_{\mu,\nu\alpha}u_{\nu,\mu\alpha}
\end{align}
and thus $\FY_{\mu\nu,\mu\nu}-\FZ=u_{\nu,\mu\alpha}u_{\mu,\nu\alpha}$. The
cancellation of so many terms is quite remarkable.

Finally \eqref{eqn:T2P} becomes
\begin{align}
\FT_2[\vu]=&\half\int\d\vr \Big\{ \T_{\mu\nu}
\lbr u_{\mu,\alpha}u_{\nu,\alpha} + u_{\alpha,\mu}u_{\nu,\alpha}
+ u_{\mu,\alpha}u_{\alpha,\nu} \rbr
\nonumber\\&
\hspace{11mm}
+ (\n/4) u_{\nu,\mu\alpha}u_{\mu,\nu\alpha} \Big\}
\\
=&-\half\int\d\vr u_{\mu}\Kh_{\mu\nu}u_{\nu}
\label{eqn:T2KHat}
\\
\Kh_{\mu\nu}=& -\frac14 \Dn\Da \n \Da\Dm
\nonumber\\&
+ \Da \T_{\mu\nu} \Da 
+ \Da \T_{\alpha\nu}\Dm + \Dn \T_{\alpha\mu} \Da.
\label{eqn:KHat2}
\end{align}
Here we used integration by parts on the derivatives of $u_{\mu}$
to obtain \eqref{eqn:T2KHat} and \eqref{eqn:KHat2}. The operator
$\Kht$ defined in \eqref{eqn:KHat2} is identical to
that defined in \eqref{eqn:KHat}.
Taking the functional derivative w.r.t. $u_{\mu}$ thus gives
\begin{align}
\vF^{1\Kin}_{\mu}=&
\frac{\delta \FT_2[\vu]}{\delta u_{\mu}}=-\Kh_{\mu\nu} u_{\nu}
\end{align}
and it is clear that \eqref{eqn:dT2u} is satisfied.

\section{Correlation energy expressions}
\label{app:Ec}

Let us first define the projected Coulomb operator
$W_{NM}=\bra{\vu_M}\Qht\ket{\vu_N}$
where $\Qht$ is defined in \eqref{eqn:FInt} or \eqref{eqn:QtDef}.
This can be written as
\begin{align}
W_{NM}=&\int\d\vr \vu_M^*(\vr)\cdot[\Qht\vu_N](\vr)
\\
=&\int\d\vr\d\vrp \vu_M^*(\vr)\cdot\ten{Q}(\vr,\vrp)\cdot\vu_N(\vrp)
\\
=&\int\frac{\d\vr\d\vrp}{|\vr-\vrp|}\DM^*(\vr)\DN(\vrp).
\end{align}
where similar equivalences hold true for
$W_{NM\lambda}=\bra{\vu_{M\lambda}}\Qht\ket{\vu_{N\lambda}}$
or for alternative forms of the Coulomb potential
(e.g. range-separated).

Using equations \eqref{eqn:QCMEig}-\eqref{eqn:QCMEigL}
and \eqref{eqn:Chi0}-\eqref{eqn:ChiL} we can write
the correlation energy as
\begin{align}
\Ec=&\half\int_0^1\d\lambda \int_0^{\infty}\frac{\d\sigma}{\pi}
\int\frac{\d\vr\d\vrp}{|\vr-\vrp|}
\nonumber\\& \times \sum_N\lbr
\frac{\DNl^*(\vr)\DNl(\vrp)}{\Omega_{N\lambda}^2+\sigma^2}
-\frac{\DN^*(\vr)\DN(\vrp)}{\Omega_{N}^2+\sigma^2} \rbr
\\
=&\half\int_0^1\d\lambda \int_0^{\infty}\frac{\d\sigma}{\pi}
%\nonumber\\&\times
\sum_N\lbr \frac{W_{NN\lambda}}{\Omega_{N\lambda}^2+\sigma^2}
- \frac{W_{NN}}{\Omega_{N}^2+\sigma^2}
\rbr
\label{eqn:EcForm1}
\\
=&\half\int_0^1\d\lambda
%\nonumber\\&
\sum_N\lbr \frac{W_{NN\lambda}}{2\Omega_{N\lambda}}
- \frac{W_{NN}}{2\Omega_{N}} \rbr.
\label{eqn:EcForm2}
\end{align}
Here the governing eigen-equations for $\vu_{N\lambda}$ are
as defined in \eqref{eqn:QCMEigL}
\begin{align}
\Omega_{N\lambda}^2\n\vu_{N\lambda}=&(\Rht+\lambda\Qht)\vu_{N\lambda}
\label{eqn:QCMEigLApp}
\end{align}
with normalisation
$\int\n(\vr)\vu_{N\lambda}^*(\vr)\cdot\vu_{M\lambda}(\vr)=\delta_{NM}$.

Following the ideas of Furche\cite{Furche2008}, we can take
the $\lambda$ derivative of $\int\d\vr\vu_{N\lambda}\cdot$\eqref{eqn:QCMEigLApp}
to work directly from \eqref{eqn:EcForm2}. Here
\begin{align}
2\Omega_{N\lambda}[\Dp{\lambda}\Omega_{N\lambda}]=&
\int\d\vr \vu_{N\lambda}^*\cdot\Qht\vu_{N\lambda}
\end{align}
where derivatives of $\vu_{N\lambda}$
can be ignored by the Hellman-Feynman theorem.
Thus
\begin{align}
[\Dp{\lambda}\Omega_{N\lambda}]=&\int\d\vr
\frac{\vu_{N\lambda}^*(\vr)[\Qht\vu_{N\lambda}](\vr)}%
{2\Omega_{N\lambda}}=\frac{W_{NN\lambda}}{2\Omega_{N\lambda}}.
\end{align}
and we can write the correlation as a sum over zero-point
energies such that
\begin{align}
\Ec=&\half\int_0^1\d\lambda
\sum_N\lbr[\Dp{\lambda}\Omega_{N\lambda}] - \frac{W_{NN}}{2\Omega_N} \rbr
\\
=&\half\sum_N\lbrs
 \Omegab_N - \Omega_N\lbr 1 + \frac{W_{NN}}{2\Omega_N^2}\rbr \rbrs
\label{eqn:EcZP}
\end{align}
where $\Omegab_N=\Omega_{N1}$. In certain systems it may make
sense to work with exchange and correlation together. Here
\begin{align}
\Exc=&\half\sum_N\lbrs \Omegab_N - \Omega_N \rbrs
- \half\int\d\vr \n(\vr)w_C(\vr)
\label{eqn:ExcZP}
\end{align}
where $w_C(\vr)=\int\d\vrp \delta(\vr-\vrp) V_C(|\vr-\vrp|)$
is like the Coulomb potential at zero distance and is
ill-defined for a true Coulomb potential $V_C(R)=1/R$.
However it becomes well-defined if we replace the Coulomb
potential by a range-separated\cite{Savin1995,Leininger1997,Gerber2005}
form eg. $V_C^{(q_{\rm RS})}(R)=\textrm{erf}(q_{\rm RS} R)/R$.

To solve for $\Omegab_N$ involves a difficult diagonalisation
of $\Rht+\Qht$ and may be best avoided.
We can use the complete and orthogonal nature of $\vu_{N\lambda}$
and $\vu_N$ to write
$\vu_{N\lambda}=\sum_K U_{NK}\vu_K$ where $\mat{U}^{\dag}\mat{U}=\mat{I}$.
Thus we can write
\begin{align}
\chi_{\lambda}(\vr,\vrp)=&-\Re\sum_{NM} X_{NM\lambda}\DN^*(\vr)\DM(\vrp)
\end{align}
where $X_{NM0}=\delta_{NM}/(\sigma^2+\Omega_N^2)$
or $\mat{X}_0=(\sigma^2+\mat{L})^{-1}$
where $L_{NM}=\delta_{NM}\Omega_N^2$. Solving for
$\chih_{\lambda}=\chih_0+\lambda\chih_0\hat{v}\chih_{\lambda}$ gives
\begin{align}
\mat{X}_{\lambda}=&\mat{X}_0 - \lambda\mat{X}_0\mat{W}\mat{X}_{\lambda},
&
\mat{X}_{\lambda}=&\frac{\mat{I}}{\sigma^2+\mat{L}+\lambda\mat{W}}.
\end{align}
This corresponds to
\begin{align}
\Ec=&\half\int_0^1\d\lambda\int_0^{\infty}\frac{\d\sigma}{\pi}
\nonumber\\&\times
\tr\lbrs\frac{\mat{W}}{\sigma^2+\mat{L}+\lambda\mat{W}}
-\frac{\mat{W}}{\sigma^2+\mat{L}} \rbrs
\end{align}
which can sometimes prove useful in real calculations

We can relate all this back to \eqref{eqn:EcZP}.
Solving the eigen-equation $\mat{V}\mat{D}_{\lambda}
=(\mat{L}+\lambda\mat{W})\mat{V}$
gives
\begin{align}
X_{MN\lambda}=&[\mat{V}(\sigma^2+\mat{D}_{\lambda})^{-1}\mat{V}^{\dag}]_{MN}
\\
=&\sum_K U^*_{KM}(\sigma^2+\Omega_{K\lambda}^2)^{-1}U_{KN}.
\end{align}
With appropriate sorting of the eigen-value/vector pairs
it is clear that the eigenvalues $D_{NN1}$ of $\mat{L}+\mat{W}$
are $\Omegab_N^2=\Omega_{N1}^2$ and $\mat{U}=\mat{V}^{\dag}$.
Thus the two expressions are equivalent.

Furthermore for $\Omega_N^2\gg \bar{W}_{N}$ where $\bar{W}_N=\sum_M |W_{NM}|$
we can solve the
eigen-problem perturbatively so that $\Omegab_N^2\aeq\Omega_N^2+W_{NN}$.
Thus $\Omegab_N=\sqrt{\Omega_N^2+W_{NN}}\aeq
\Omega_N(1 + \frac{W_{NN}}{2\Omega_N^2} - \frac{W_{NN}^2}{8\Omega_N^4})$.
We can speed calculation and improve numerical stability
by choosing an $N^*$ above which we use the approximation.
Setting $\beta_N=\frac{W_{NN}}{2\Omega_N^2}$ we then find
\begin{align}
\Ec\aeq &\half\sum_{N=0}^{N^*}
\lbrs \Omegab_N-\Omega_N\lbr 1+\beta_N\rbr \rbrs
%\nonumber\\&
-\sum_{N>N^*}\frac{\Omega_N\beta_N^2}{4}.
\end{align}
Such a perturbative approach will be almost guaranteed convergent,
and with a fairly small $\Omega_{N^*}$ if range-separation is used.

\section{One-Electron governing operator}
\label{app:OneEl}

As discussed in \sref{sec:OneEl}
[equations \eqref{eqn:OneElPhi}-\eqref{eqn:OEQCM1}]
we can obtain the exact linear perturbation solutions in a
one-electron system via a solution of
\begin{align}
\omega^2\vd=&\RhtOE\vd
\end{align}
where the associated change in density is
$n^1=-\vnabla\cdot\psi\vdOE$. Since this is an identical
problem to finding the QCM solutions via
\begin{align}
\omega^2\vd=&\Rhdt\vd
\end{align}
it is clear that if $\RhtOE=\Rhdt$ then the QCM is exact
for one-electron systems.

For a one-electron system we set $V=V^{\KS}-\epsilon_0$
such that the groundstate occupied orbital $\psi$
is a solution of $[-\half\nabla^2+V]\psi=0$.
Equations \eqref{eqn:OneElOrb}
thus become $\Ra=\Da\log|\psi|$ and $S=2V$
and thus $\hh=-\half(\nabla^2-S)$.
The former can be used to derive the following
\begin{align*}
[\Dm\Rn]=&[\Dn\Rm]=-2t_{\mu\nu}
\\
[\nabla^2\Rm]=&-2[\Da t_{\alpha\mu}]=S_{,\mu} + 4\Ra t_{\alpha\mu}
\end{align*}
where these identities will be used throughout this appendix.

Comparing the final term of \eqref{eqn:OEQCM1} with \eqref{eqn:RHat}
we wish to swap $(\Dm-\Rm)$ and $(\Dn+\Rn)$ across sides. This
can be done by repeatedly using
\begin{align*}
\Dm f=&f \Dm + [\Dm f]
\\
\nabla^2 f =& f \nabla^2 + [\nabla^2 f] + 2[\Da f]\Da
\\
=& f \nabla^2 - [\nabla^2 f] + 2\Da[\Da f]
\\
f \nabla^2 =& \nabla^2 f + [\nabla^2 f] - 2\Da[\Da f]
\\
=& \nabla^2 f - [\nabla^2 f] - 2[\Da f]\Da
\end{align*}
in the appropriate places.

We first address the terms involving $S$. These are
\begin{align*}
\Dm S \Dn=&\Dn S \Dm + S_{,\mu}\Dn - \Dm S_{,\nu} + S_{,\mu\nu}
\\
\Dm S \Rn =& \Rn S \Dm + [\Dm \Rn S]
\\
-\Rm S \Dn =& -\Dn S \Rn + [\Dn \Rm S]
\end{align*}
and thus
\begin{align}
\RhOE_{\mu\nu-S}=&(\Dm-\Rm)S(\Dn+\Rn)
\\=&
(\Dn+\Rn)S(\Dm-\Rm)
\nonumber\\&
+ S_{,\mu}\Dn - \Dm S_{,\nu} + S_{,\mu\nu}
\nonumber\\&
 + [\Dm \Rn S] + [\Dn \Rm S].
\label{eqn:OES}
\end{align}

The terms involving the Laplacian are a little
more difficult to deal with. Here
\begin{align*}
\Dm\nabla^2\Rn
=&\Rn\nabla^2\Dm - 2\nabla^2t_{\mu\nu}
\nonumber\\&
- S_{,\nu} \Dm + 4\Ra t_{\alpha\nu} \Dm
- 4\Da t_{\alpha\nu}\Dm
\\
-\Rm\nabla^2\Dn
=&-\Dn\nabla^2\Rm - 2t_{\mu\nu}\nabla^2
\nonumber\\&
+ \Dn S_{,\mu} + 4\Dn t_{\alpha\mu} \Ra
- 4\Dn t_{\alpha\mu}\Da
\end{align*}
and
\begin{align*}
\Rm\nabla^2\Rn
=&(\Da\Rm - [\Da\Rm])(\Rn\Da + [\Da\Rn])
%\\
%=&\Da\Rn\Rm\Da - [\Da\Rm][\Da\Rn]
%\nonumber\\&
%- [\Da\Rm]\Rn\Da + \Da\Rm[\Da\Rn]
\\
=&(\Rn\Da+[\Da\Rn])(\Da\Rm-[\Da\Rm])
\nonumber\\&
 - [\Da\Rm]\Rn\Da + \Da\Rm[\Da\Rn] - [\Da\Rm][\Da\Rn]
%\\
%=&\Rn\nabla^2\Rm-2[\Da\Rn][\Da\Rm]
%\nonumber\\&
%+ [\Da\Rn]\Da\Rm - \Rn\Da[\Da\Rm]
%\nonumber\\&
%- [\Da\Rm]\Rn\Da + \Da\Rm[\Da\Rn]
\\
=&\Rn\nabla^2\Rm-2[\Da\Rn][\Da\Rm]
\nonumber\\&
- \Rn[\nabla^2\Rm] - [\nabla^2\Rn]\Rm
\nonumber\\&
- 2\Rn[\Da\Rm]\Da + 2\Da[\Da\Rn]\Rm
\\
=&\Rn\nabla^2\Rm-2[\Da\Rn][\Da\Rm]
\nonumber\\&
- \Rn S_{,\mu} - \Rm S_{,\nu}
- 4\Rn t_{\alpha\mu}\Ra - 4\Ra t_{\alpha\nu}\Rm
\nonumber\\&
+ 4\Rn t_{\alpha\mu}\Da - 4\Da t_{\alpha\nu}\Rm.
\end{align*}
Combining the above lets us write
\begin{align}
\RhOE_{\mu\nu-L}
=&(\Dm-\Rm)\nabla^2(\Dn+\Rn)
%\\
%=&(\Dn+\Rn)\nabla^2(\Dm-\Rm)
%\nonumber\\&
% - 2\nabla^2t_{\mu\nu}
% - 2t_{\mu\nu}\nabla^2
%\nonumber\\&
%- S_{,\nu} \Dm + 4\Ra t_{\alpha\nu} \Dm
%- 4\Da t_{\alpha\nu}\Dm
%\nonumber\\&
%+ \Dn S_{,\mu} + 4\Dn t_{\alpha\mu} \Ra
%- 4\Dn t_{\alpha\mu}\Da
%\nonumber\\&
%+2[\Da\Rn][\Da\Rm]
%\nonumber\\&
%+ \Rn S_{,\mu} + \Rm S_{,\nu}
%+ 4\Rn t_{\alpha\mu}\Ra + 4\Ra t_{\alpha\nu}\Rm
%\nonumber\\&
%- 4\Rn t_{\alpha\mu}\Da + 4\Da t_{\alpha\nu}\Rm
\\
=&(\Dn+\Rn)\nabla^2(\Dm-\Rm)
-4\Da t_{\mu\nu} \Da
\nonumber\\&
- 4(\Da+\Ra) t_{\alpha\nu}(\Dm-\Rm)
\nonumber\\&
- 4(\Dn+\Rn) t_{\alpha\mu}(\Da-\Ra)
\nonumber\\&
+2[\Da\Rn][\Da\Rm] - 2[\nabla^2 t_{\mu\nu}]
\nonumber\\&
+ \Dn S_{,\mu}- S_{,\nu} \Dm
+ \Rn S_{,\mu} + \Rm S_{,\nu}
\label{eqn:OELap}
\end{align}

We can now use \eqref{eqn:OELap} and \eqref{eqn:OES}
together to show
\begin{align}
4\RhOE_{\mu\nu}=&\RhOE_{\mu\nu-L}-\RhOE_{\mu\nu-S}
%=&(\Dm-\Rm)(\nabla^2-S)(\Dn+\Rn)
%\\
%=&(\Dn+\Rn)(\nabla^2-S)(\Dm-\Rm)
%\nonumber\\&
%-4\Da t_{\mu\nu} \Da
%- [\Dm \Rn S] - [\Dn \Rm S]
%\nonumber\\&
%- 4(\Da+\Ra) t_{\alpha\nu}(\Dm-\Rm)
%\nonumber\\&
%- 4(\Dn+\Rn) t_{\alpha\mu}(\Da-\Ra)
%\nonumber\\&
%+ S_{,\mu\nu}
%+2[\Da\Rn][\Da\Rm] - 2[\nabla^2 t_{\mu\nu}]
%\nonumber\\&
%+ \Rn S_{,\mu} + \Rm S_{,\nu}.
\\
=&(\Dn+\Rn)(\nabla^2-S)(\Dm-\Rm)
\nonumber\\&
-4(\Da+\Ra)t_{\mu\nu}(\Da-\Ra)
\nonumber\\&
- 4(\Da+\Ra) t_{\alpha\nu}(\Dm-\Rm)
\nonumber\\&
- 4(\Dn+\Rn) t_{\alpha\mu}(\Da-\Ra)
\nonumber\\&
+ S_{,\mu\nu}+2[\Da\Rn][\Da\Rm]
\nonumber\\&
- 2[\nabla^2 t_{\mu\nu}]
- 4[\Ra\Da t_{\mu\nu}]
\label{eqn:OEPen}
\end{align}
where we cancel most terms via
\begin{align*}
\Da t_{\mu\nu} \Da=&(\Da+\Ra) t_{\mu\nu} (\Da-\Ra)
\nonumber\\&
 + [\Ra\Da t_{\mu\nu}] +  S t_{\mu\nu}
\\
[\Dm\Rn S]=&-2S t_{\mu\nu} + [\Rn S_{,\mu}]
\\
[\Dn\Rm S]=&-2S t_{\mu\nu} + [\Rm S_{,\nu}].
\end{align*}

The operator terms are now the same as those of \eqref{eqn:Rhd}.
The remaining constant is
\begin{align*}
4K_{\mu\nu}=&S_{,\mu\nu}+2[\Da\Rn][\Da\Rm] - 2[\nabla^2 t_{\mu\nu}]
- 4[\Ra\Da t_{\mu\nu}]
\\
=&S_{,\mu\nu}+2[\Da\Rn][\Da\Rm] + S_{,\mu\nu}
\nonumber\\&
-2[\Dn\Ra\Dm\Ra]+2[\Ra\Da\Dm\Rn]
\\
=&2S_{,\mu\nu}=4V_{,\mu\nu}
\end{align*}
where we have used
$[\nabla^2 t_{\mu\nu}]=-\half S_{,\mu\nu} + [\Dn\Ra][\Dm\Ra]+[\Ra\Dn\Dm\Ra]$
and $[\Da\Rn]=[\Dn\Ra]$ to arrive at the final form.

Thus \eqref{eqn:OEPen} becomes
\begin{align}
\RhOE_{\mu\nu}=&
V_{,\mu\nu} + \frac14(\Dn+\Rn)(\nabla^2-S)(\Dm-\Rm)
\nonumber\\&
-(\Da+\Ra)t_{\mu\nu}(\Da-\Ra)
\nonumber\\&
-(\Da+\Ra) t_{\alpha\nu}(\Dm-\Rm)
\nonumber\\&
-(\Dn+\Rn) t_{\alpha\mu}(\Da-\Ra)
\end{align}
which is identical to $\Rhd_{\mu\nu}$ and through it
$\Rh_{\mu\nu}$ via \eqref{eqn:RHatd}.

Finally, since $\psi^1$ and through it $\phi^1$ and $n^1$ can be calcuted
from $\vd$ it is clear that a solution to the QCM is identical to a
linearised solution of the Schr\"odinger equation for a one-electron
systen, and vice versa.

\section{Scalar QCM operator in the KS basis}
\label{app:KS}
Inserting \eqref{eqn:Rhd} [and setting $(\nabla^2-S)=(\Da+\Ra)(\Da-\Ra)$]
into \eqref{eqn:RhDefKS} and using
integration by parts we find
\begin{align}
R_{jk}=&\int\d\vr [\Dfm\psi_j^*]V_{,\mu\nu}[\Dfn\psi_k]
\nonumber\\&
+ \frac14 \int\d\vr [\Dfa\Dfm\Dfn\psi_j^*]
[\Dfa\Dfn\Dfm\psi_k]
\nonumber\\&
+ \int\d\vr [\Dfa\Dfm\psi_j^*]
t_{\mu\nu}[\Dfa\Dfn\psi_k]
\nonumber\\&
+ \int\d\vr [\Dfa\Dfm\psi_j^*]
t_{\alpha\nu}[\Dfm\Dfn\psi_k]
\nonumber\\&
+ \int\d\vr [\Dfn\Dfm\psi_j^*]
t_{\alpha\mu}[\Dfa\Dfn\psi_k]
\label{eqn:RjkA}
\end{align}
where we have used the shorthand $\Dfa=\Da-\Ra$.

It follows from $\Rv=\half\nabla\log\n$ that $[\Dm\Rn]=[\Dn\Rm]$
and thus $\Dm\Rn + \Rm\Dn=\Rn\Dm + \Dn\Rm$. Therefore
$\Dfm\Dfn=\Dfn\Dfm$ and we can
simplify \eqref{eqn:RjkA} (noting that we can also swap
Greek indices as they are summed over) to
\begin{align}
R_{jk}=&\int\d\vr [\Dfm\psi_j^*]V_{,\mu\nu}[\Dfn\psi_k]
\nonumber\\&
+ \frac14 \int\d\vr [\Dfa\Dfm\Dfn\psi_j^*]
[\Dfa\Dfm\Dfn\psi_k]
\nonumber\\&
+ 3\int\d\vr [\Dfa\Dfm\psi_j^*]
t_{\mu\nu}[\Dfa\Dfn\psi_k].
\end{align}

\section*{References}
\bibliography{vanDerWaals,ACFDT,DFT,Wannier,Misc,Experiment,Hybrid,%
CM-RPA,QMGeneral}

\end{document}